\documentclass[oldversion]{aa}
\usepackage{natbib}
\usepackage{graphicx}
\usepackage{longtable}
\usepackage{txfonts}
\usepackage{ifpdf}

\setlongtables

\newcommand{\logten}{\ensuremath{\log_{10}}}
\newcommand{\remnantabbr}{remnant}
\newcommand\refereebf[1]{{\bf #1}}
\renewcommand\refereebf[1]{{#1}}
\newcommand\beq{\begin{equation}}
\newcommand\eeq{\end{equation}}
\providecommand{\avg}[1]{\ensuremath{\left< #1 \right >}}

\ifpdf
   \newcommand{\imgangle}{0}
\else
   \newcommand{\imgangle}{270}
\fi

\titlerunning{A grid of low-mass collisions}
\title{Evolution of stellar collision products in open clusters. II. 
A grid of low-mass collisions}
\author{
   Evert Glebbeek \and
   Onno R. Pols
}

\institute{
Sterrekundig Instituut Utrecht, Postbus 80000, 3508 TA Utrecht, The Netherlands.
}

\abstract{
In a companion paper we studied the detailed evolution of stellar collision
products that occurred in
an $N$-body simulation of the old open cluster M67 and compared our
detailed models to simple prescriptions.
In this paper we extend this work by studying the evolution of the
collision products in open clusters as a function of mass and age of the progenitor stars.

We calculated a grid of head-on collisions covering the section of
parameter space relevant for collisions in open clusters.
We create detailed models of the merger remnants using
an entropy-sorting algorithm and follow their subsequent evolution during
the initial contraction phase, through the main sequence and up to the
giant branch with our detailed stellar evolution code.
We compare the location of our models in a colour-magnitude diagram to the observed
blue straggler population of the old open clusters M67 and NGC 188 and find
that they cover the observed blue straggler region of both clusters. For M67, collisions need to have
taken place recently.
Differences between the evolution tracks of the collision products and normal
main sequence stars can be understood quantitatively using a simple
analytic model. We present an analytic recipe that can be used in an
$N$-body code to transform a precomputed evolution track 
for 
a normal star into an evolution track for a collision product.
}

\date{Today}

\bibpunct{(}{)}{;}{a}{}{,} 

\newlength{\timeswidth}
\newlength{\pluswidth}
\newcommand{\plustimes}{\ensuremath{%
\settowidth{\timeswidth}{$\times$}%
\settowidth{\pluswidth}{$+$}%
\addtolength{\timeswidth}{\pluswidth}%
+\hspace{-.5\timeswidth}\times%
}}

\begin{document}

\maketitle

\section{Introduction}


Star clusters are important laboratories for a wide range of
astrophysical processes. It has become clear that the evolution of a
star cluster is driven by the complex interplay between stellar
dynamics and stellar and binary evolution
\citep{1999A&A...348..117P,article:hurley_first_m67}. Physical
collisions between stars in the dense cluster core play a pivotal role
in this interaction. The products of stellar collisions between
main-sequence stars potentially stand out as blue stragglers in a
colour-magnitude diagram. The blue straggler population of a cluster
can therefore be used to study its dynamical history. Since stellar
collision products generally have a very different thermal and
chemical structure than normal main-sequence stars
\citep{lombardi_collisions}, detailed calculations of their evolution
are required
\citep{article:sills_evcolprod1,sills_evcolprod2}.

In a previous paper \citep{paperI} we have developed an efficient
procedure to import detailed models of stellar collision products into
a fully implicit stellar evolution code and to evolve the remnants
well beyond the main sequence. Our work is similar to that of
\citet{article:sills_evcolprod1} but our code is faster and much more
robust, allowing for the first time a systematic study of stellar
merger remnants and an exploration of the parameter space. 
\citet{paperI} studied the evolution of collision remnants that
occurred in the $N$-body simulation of M67 by
\citet{article:hurley_m67} and compared these with the evolution
tracks predicted by the prescription of \citet{article:hurley_bse} as
well as tracks of fully mixed versions of the remnants. We found that
our merger remnants have shorter main-sequence lifetimes than
predicted by either the \citet{article:hurley_bse} prescription or the
fully mixed models. Our models are also brighter than normal stars of
the same mass, but not as blue as fully mixed models.

In this work we explore the parameter space for collisions between
low-mass main sequence stars relevant for blue straggler formation in
old open clusters (such as M67) and place the findings of
\citet{paperI} in context. 
\refereebf{A large grid of stellar collisions covering this parameter space
was calculated by \citet{2005MNRAS.358.1133F}. They mainly focussed on
collisions with high impact velocity suitable for the galactic centre and
their simulations had insufficient resolution to resolve mixing due to the
collision. Our work complements theirs by investigating the mixing as well as
the long term evolution of the merger remnants.}
We describe the detailed structure and
evolution of the collision products and discuss their dependence on
the collision parameters (masses and time of collision) in
\S\ref{sec:properties_and_tools}. We compare the results of our
models with observations of blue stragglers in the open clusters M67
and NGC188 in \S\ref{sec:comparison}. Finally, we show how the global
properties of the collision products (luminosity, radius and lifetime)
can be described by a simple analytic prescription that can be
included in a parametric model of stellar collisions and in $N$-body
simulations (\S\ref{sec:recipe}). We discuss our results in
\S\ref{sec:discussion}.

\section{Initial conditions and parameters of the collisions}
\label{sec:initial_conditions}

The structure of a merger remnant depends on the structure of the two parent 
stars, the impact parameter of the collision and the relative velocity
at infinity. The structure of the parent stars depends on their masses,
chemical composition (characterised by their heavy-element content $Z$) and
their evolutionary stage. In this work we have restricted ourselves to a
section of this 8-dimensional parameter space.

First of all we assume that the two stars involved in the collision are coeval
in the sense that both are at the zero-age main sequence (ZAMS) at $t=0$
and have the same initial composition, which is a reasonable assumption for
stars in clusters. 
\refereebf{Placing $t=0$ at an earlier stage, \emph{e.g.} the onset of deuterium burning rather than the
ZAMS, will not greatly affect the outcome of our calculations because
appreciable composition gradients only build up on the main sequence and we
consider a fairly small mass range.}
We restrict ourselves to collisions between
low-mass main sequence stars (by which here we mean that their combined
mass does not exceed $2.4 M_\odot$, see below) and consider only
head-on collisions (i.e., with impact parameter
$b=0$), in essence meaning that we ignore the effect of rotation in the
collision product. This is despite the fact that rotation can be of
considerable importance for the structure and evolution of the collision
product: even for collisions with a small impact parameter the remnant has
sufficient angular momentum that a main sequence star of the same mass
would need to rotate faster than its breakup
rate \citep{lombardi_collisions,article:sills_evcolprod1}.
The physical mechanism and timescale on which the collision products lose
their angular momentum are unclear. Methods to incorporate the effect of 
rotation in a one dimensional stellar evolution code treat rotation as a perturbation
to a non-rotating stellar model
\citep{endal_sofia_rotation,pinsonneault_rotation,zahn_rotation,heger_rotation}.
\refereebf{These methods are accurate for rotation rates less than about
60\% of the critical (Keplerian) value \citep{2004A&A...425..217Y} and it
is not clear how to model stars that are even closer to critical rotation.}
We plan to investigate this problem in future research.

\begin{table*}
\caption{Values used for the different grid parameters. The grid spacing is
listed between parentheses. Note that the grids labelled \emph{M67} and
\emph{NGC188} overlap in part of their age range. We have calculated models
for two extra mass ratios for a limited mass range in the \emph{M67} grid
to more clearly resolve trends at high and low mass ratios.
}\label{tab:grid_param}
\begin{tabular}{lllll}
\hline\hline
Name & $Z$ & $t [\,\mathrm{Myr}]$ & $M [M_\odot]$ & $q$ \\
\hline
M67 & $0.02$ & $2800$, $3100$, $3400$, $3700$ & $1.5$--$2.4$ (0.1) & $0.4$--$1.0$ (0.2)\\

& & & $1.5$--$2.0$ (0.1) & $0.5$, $0.9$\\

NGC188 & $0.02$ & $3400, 3700, 4200, 4700, 5200, 5700$ & $1.2$, $1.3$, $1.4$, $1.6$, $1.8$, $2.0$, $2.2$ & $0.4$--$1.0$ (0.2)\\


GC & $0.001$ & 8000--12500 (1500) & $0.8$--$1.3$ (0.1) & $0.4$--$1.0$ (0.2)\\
\hline
\end{tabular}
\end{table*}

This leaves us with four parameters: the composition, the time of collision $t$ and the
masses of the two progenitors. Borrowing nomenclature from the field of
binary evolution, we refer to the more massive progenitor as the primary
and write its mass as $M_1$. Similarly we refer to the less massive
progenitor as the secondary and denote its mass by $M_2$. With these
definitions we can introduce the total initial mass $M=M_1+M_2$ and the
mass ratio $q=M_2/M_1$. To calculate a grid of models we have used the
independent parameters $Z$, $t$, $M$ and $q$.
We have calculated several grids covering different parts of this parameter
space, as listed in Table \ref{tab:grid_param}. The grid labelled
\emph{M67} covers the parameter space sampled by the $N$-body simulation of
\citet{article:hurley_m67} and is relevant for old open clusters like M67.
An extension to the \emph{M67} grid is the parameter set labelled \emph{NGC188},
which covers the parameter range of interest for the somewhat older open
cluster NGC188.
Both these grids use an assumed heavy-element content $Z=0.02$ and an
initial hydrogen abundance $X=0.70$. The 
total initial masses for the collision products were chosen such that the
lower limit is just above the present-day turnoff mass and the upper limit
is roughly twice the turnoff mass ($1.18 M_\odot$ for NGC188, $1.29
M_\odot$ for M67); see \S \ref{sec:comparison}.

We have also computed a grid (labelled \emph{GC}) with $Z=0.001$ and
$X=0.757$ that is suitable for
comparison with globular clusters. In the presentation of our results we
will mostly focus on the \emph{M67} and \emph{NGC188} grids but unless
otherwise indicated our results apply to the $Z=0.001$ grid as well.

\section{Tools}
\label{sec:tools}

The method we use in our study has been described in detail in a
separate paper \citep{paperI} (hereafter Paper I), so we refer the
interested reader to that paper for more details about the procedure of
calculating the models and provide just a brief summary here.

For each collision we evolve models of the progenitor stars to the time of
collision and then merge the two stars.  The structure of the merger
remnants is calculated using the Make Me A Star (MMAS) code developed by
\citet{article:lombardi_mmas}. This code uses the realisation that in low
mass stars the quantity
\begin{equation}
A = \frac{P}{\rho^{5/3}}
\end{equation}
increases outward. The quantity $A$ is closely related to the entropy and
the tendency of $A$ to increase reflects the tendency of low entropy
material to sink to the centre of the collision product. For this reason
the procedure by which the remnant profile is constructed is known as
\emph{entropy sorting}.

Our evolution code is a version of the STARS code developed by \citet{article:eggleton_evlowmass} and updated by others \citep{article:pols_approxphys}.
Nuclear reaction rates are from \citet{article:caughlan_and_fowler_reactionrates}
and \citet{article:caughlan_reactionrates}
and we use opacities from \citet{article:opal1992} and \citet{article:alexander_ferguson_lowtopac}. The assumed 
heavy-element composition is scaled to solar abundances 
\citep{article:anders_and_grevesse_abund}.
Chemical mixing due to convection
\citep{article:bohm-vitense_convection,article:eggleton_mixingproc} and
thermohaline mixing \citep{article:kippenhahn_thermohalinemixing} is taken
into account. 
All models are computed with a mixing-length ratio $l/H_P=2.0$. As in
Paper~I, we have neglected convective overshooting in all models in this
work.

\section{Properties and structure of collision products}
\label{sec:properties_and_tools}

In this section we present the structure and evolution of the merger
remnants. We will discuss how the structure of the parent stars affects the
structure of the remnant. Details for all our collision models are listed in Table
\ref{tab:model_table} for $Z=0.02$ models and in Table
\ref{tab:model_z001_table} for $Z=0.001$ models. \refereebf{Both these
tables are available in the Online version of this paper.}

\subsection{Mass loss from the collision}\label{sec:massloss}
A fraction $\phi$ of the total mass $M_1+M_2$ of the progenitor stars is
lost during the collision. The fraction $\phi$ is, for central collisions, 
estimated in MMAS as
\beq\label{eqn:massloss}
\phi = C \frac{q}{(1+q)^2}\frac{R_{1,0.86} + R_{2,0.86}}{R_{1,0.5} + R_{2,0.5}},
\eeq
\citep{article:lombardi_mmas}, where $R_{n,0.86}$ and $R_{n,0.5}$ are the radii
containing 86\% and 50\% of the mass of parent star $n$ (1 or 2). The 
constant $C=0.157$ is determined by calibration to SPH calculations.
Typical values of $\phi$ are in the order of a few percent. A reasonable
approximation over the range of our grid is (see Table
\ref{tab:model_table})
\beq\label{eqn:massloss_approx}
\phi = 0.3 \frac{q}{(1+q)^2}.
\eeq

Material lost from the stars in the ejecta originates from the envelopes of the
parent stars and has the composition of the parent star envelope.

\subsection{Structure of the progenitor stars}
We can distinguish four different collision scenarios between main sequence
stars based on the structure of the progenitor stars and the collision
product.
Low mass main sequence stars ($\lesssim 1.2 M_\odot$) have radiative cores 
while main sequence stars of higher mass have a convective core. This
leads to four possible scenarios for a collision:

\begin{enumerate}
\item Convective + Convective $\to$ Convective
\item Convective + Radiative $\to$ Convective
\item Radiative + Radiative $\to$ Convective
\item Radiative + Radiative $\to$ Radiative
\end{enumerate}

The composition profile of a collision remnant is determined by the entropy
profiles of the progenitor stars, which is in turn determined by their masses and
their age. If a star has a convective core the material from the core will
have a constant entropy, which directly affects the composition profile in
the remnant and its subsequent evolution. In a star with a radiative core
the entropy increases outward.

The first case will generate remnants with masses $> 2.4 M_\odot$ and will
be the topic of a future paper. The second case occurs in our grid for the
higher mass collision products ($M>1.7 M_\odot$) at extreme mass ratios
($q>0.5$) and is relevant for clusters where the turnoff mass is larger
than $1.2 M_\odot$.
The third case makes up the majority of the collision products in our grid and 
is relevant for old open clusters as well as globular clusters. The main
difference with case 2 is that in case 2 the core of the primary has an
almost flat entropy profile whereas in case 3 the entropy profile is
smoother.
The final case is only relevant for clusters where the turnoff mass is below
$1.2 M_\odot$. In younger clusters the collision products of these collisions
would be hidden among normal main sequence stars and not stand out as blue
stragglers. The progenitor stars are essentially unevolved at the time of
collision and the outcome of such a merger is a new unevolved main sequence
star.

\subsection{Composition profile}\label{sec:composition_profiles}
\begin{figure}
\includegraphics[angle=0,width=0.5\textwidth]{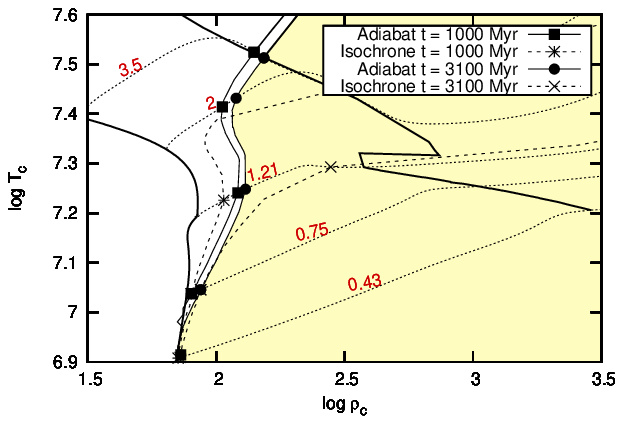}
\caption{Evolution tracks of low mass stars in the $\log \rho_\mathrm{c}$--
$\log T_\mathrm{c}$ plane. Bold lines indicate the locations of the ZAMS (left)
and TAMS (right). 
Dotted lines are evolution tracks, labelled with the ZAMS mass.
The solid lines marked with $\blacksquare$, {\Large$\bullet$}
are lines of constant entropy across different stellar models, the dashed lines
marked with $\plustimes, \times$ are isochrones. The shaded region indicates
the region where the central entropy is lower than that of a $0.75 M_\odot$ star
at 3100 Myr.}
\label{fig:rhoc_tc_tracks}
\end{figure}

During a collision material from one parent star is mixed with material 
of the same entropy from the other star (since it ends up in the same location in the
collision product). However, if one of the two stars has a convective core the 
material from the core can retain its identity because of the nearly flat
entropy profile. This can lead to a very steep composition gradient in the remnant.

On the ZAMS lower mass stars have a lower central entropy than higher mass
stars (their cores are denser), which means that in a collision between
unevolved stars material from the lower mass star will sink to the centre of 
the collision product.
As a result of stellar evolution, the entropy in the core will decrease (the 
core becomes more compact). Since more massive stars evolve more quickly, the
central entropy will decrease more rapidly in massive stars than it will in
low mass stars.

Figure \ref{fig:rhoc_tc_tracks} shows a number of evolution tracks in the
$\log \rho_\mathrm{c}$--$\log T_\mathrm{c}$ plane. The location of the ZAMS
is indicated by a thick solid line on the left and the location of the
terminal age main sequence (TAMS)
by a thick solid line on the right. Evolution tracks are dotted and
labelled with the mass of the star and two isochrones are shown, one for
$1000 \,\mathrm{Myr}$ and one for $3100 \,\mathrm{Myr}$. 
Two lines of constant entropy across different stellar models are also
drawn for different ages. The shaded region indicates the location
where the entropy is lower than in the core of a $0.75 M_\odot$ star at 
$t=3100 \,\mathrm{Myr}$. The $t=3100 \,\mathrm{Myr}$ isochrone lies almost 
completely within this region, indicating that in a collision with a more 
massive star at this age the core of the primary sinks to the centre of the 
collision product. 
Conversely, part of the $t=1000 \,\mathrm{Myr}$ isochrone lies to the
left of the line of constant entropy at that age, indicating that in a
collision with a more massive star, up to $\approx 1.9 M_\odot$,
the secondary star will sink to the centre.

\begin{figure}
\includegraphics[angle=\imgangle,width=0.5\textwidth]{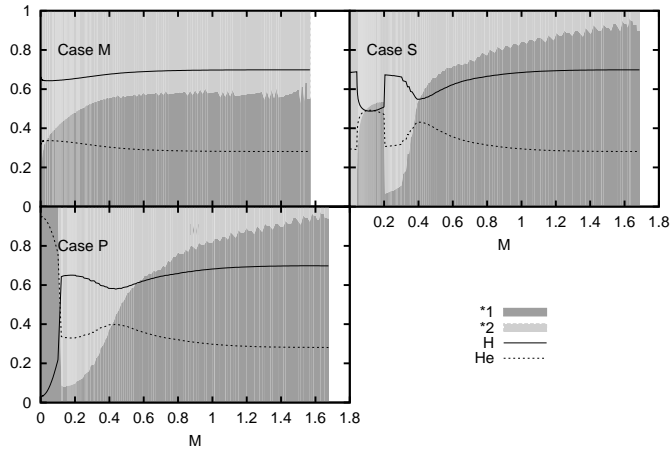}
\caption{Diagram showing the distribution of material from the parent stars
within three of the collision products: $t=2800, q = 0.8, M=1.7$
(corresponding to case \emph{M}, see text) and $t=2800, q = 0.4, M=1.8$
(case \emph{S}) and $t=3700, q=0.4, M=1.8$ (case \emph{P}) for
the bottom row. The dark shading indicates the fraction
of material from the primary while the lighter region shows the fraction of
material from the secondary. Overplotted are the abundances by mass fraction of
hydrogen (solid) and helium (dotted).}
\label{fig:mixing_example}
\end{figure}

For the composition profile in the collision product we can roughly distinguish
three cases, which we will denoted by `M', `S' and `P' \refereebf{for
`mixed core', `secondary core' and `primary core' respectively}. These are
illustrated in Figure \ref{fig:mixing_example}. The part of the
parameter space in the grid for which each of these cases occurs is illustrated in
Figure \ref{fig:mixing_cases} and listed in Table \ref{tab:model_table}.

\subsubsection{Case M}
If the mass difference between the two colliding stars is small (the mass ratio
is close to 1), the colliding stars have very similar entropy and composition 
profiles. The composition profile of the collision product will resemble 
a stretched version of the composition profile in the progenitor stars.
This situation is illustrated by the top left panel in Figure
\ref{fig:mixing_example}.
There can be a small molecular weight inversion near the centre if most of the 
material in the core comes from the secondary star. This case is indicated
by $\boxdot$ in Figure \ref{fig:mixing_cases}. In Figure \ref{fig:rhoc_tc_tracks}
this means that the two models are on or close to the line of constant
entropy.


\subsubsection{Case P}
If the primary is sufficiently evolved that its core entropy is lower than
that of the secondary, the core of the primary ends up in 
the centre of the collision product. In Figure \ref{fig:rhoc_tc_tracks} this
means that the primary is in the shaded region of the diagram.
In this case the core of the merger remnant has a reduced hydrogen content with a very steep
increase in hydrogen abundance at its edge and possibly a molecular 
weight inversion further out as well. A hydrogen burning shell can form at the 
edge of the core (see \S \ref{sec:mixing}). This is the situation in the lower left panel of Figure
\ref{fig:mixing_example} and is marked in Figure \ref{fig:mixing_cases}
with $\triangle$.
An extreme example of this case occurs when the primary is at the
end of the main sequence. The hydrogen depleted core of the primary 
sinks to the centre of the collision product, producing a new star with a 
hydrogen depleted core. In other words, collisions involving turnoff stars 
produce turnoff stars, in agreement with \citet{article:sills_evcolprod1}.

\subsubsection{Case S}
\begin{figure}
\includegraphics[width=0.5\textwidth]{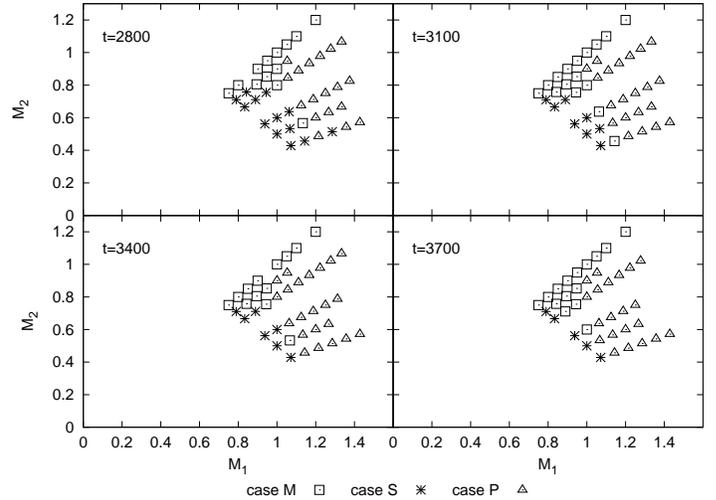}
\caption{Illustration of the different cases for the composition profiles
in the collision products and the grid parameters for which they occur.}
\label{fig:mixing_cases}
\end{figure}

If the entropy in the core of the primary is larger than the entropy in 
the core of the secondary the secondary will sink to the centre of
the collision product.  In terms of Figure \ref{fig:rhoc_tc_tracks} the
primary is to the left of the shaded region. The core of the collision product
will be hydrogen-rich, with a helium rich layer on top of it, leading to a
pronounced molecular weight inversion. This is the upper left
situation in Figure \ref{fig:mixing_example} (a $\plustimes$ in Figure
\ref{fig:mixing_cases}). 
This case will appear for collisions between main sequence stars when the mass ratio of
the colliding stars is small ($\sim 0.4$). If the mass ratio is even smaller ($\lesssim 0.2$), this will happen even after the primary has already evolved off the main sequence. 

\subsection{Reignition of hydrogen and mixing}\label{sec:mixing}
Initially there is little or no nuclear burning in the core of the collision
products and nuclear burning is unimportant as an energy source during the contraction 
to the main sequence.


Because the core of the collision product has the entropy of a lower mass star
it lies on a lower adiabat than the core of a main sequence star of the same mass and
composition. In practice, this means that the collision product is initially
to the right of its main sequence position in the $\rho_\mathrm{c}$
-- $T_\mathrm{c}$ plane, at higher central densities. In order for the star to 
come into thermal equilibrium the entropy in the core needs to increase. The core 
will first expand and then heat up (see Paper I).
Once hydrogen burning takes over as the main energy source the evolution of 
the collision product proceeds analogously to that of a normal main sequence
star but with an abnormal composition profile (again see Paper I).

\begin{figure}
\includegraphics[angle=0,width=0.5\textwidth]{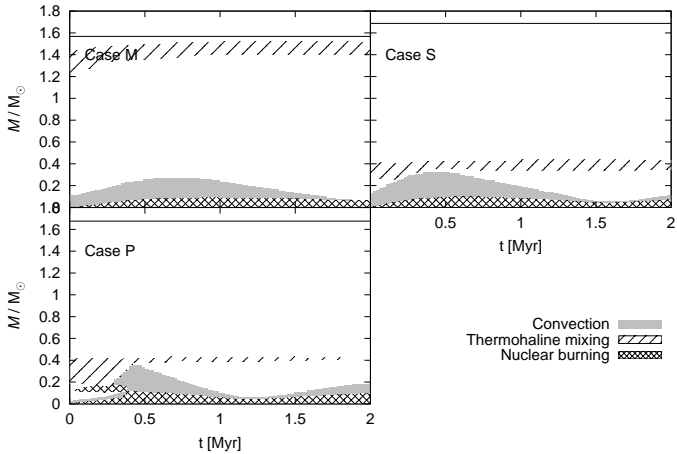}
\caption{Kippenhahn diagrams showing the mixing regions during the contraction
phase for the same collision products as shown in Figure \ref{fig:mixing_example}.}
\label{fig:kippenhahn}
\end{figure}

If there is a steep composition gradient at the edge of the core (when one of
the two stars had a convective core), it is possible for the nuclear energy
generation rate to peak at this location due to the composition dependence
of the 
reaction rate. In effect, hydrogen will burn in the core as well as in a shell
outside the core. This hydrogen burning shell is a transient feature that is
destroyed when mixing reduces the steepness of the composition gradient.
The hydrogen burning shell is important because it can drive a
convection zone that mixes the layers above it.
Examples of the mixing regions during the contraction phase in each of the three cases in Figure \ref{fig:mixing_example} are shown in Figure \ref{fig:kippenhahn}.

Because the material outside the core can be hydrogen-rich, convection can
bring extra hydrogen to the core, effectively rejuvenating the star. How
much hydrogen is mixed into the core depends on mixing processes in the
layer outside the core.

Helium-rich material on top of a layer of hydrogen-rich material causes a
molecular weight inversion, which gives rise to thermohaline mixing
\citep{article:kippenhahn_thermohalinemixing}. The material with higher
molecular weight is supported by thermal buoyancy. Diffusion of heat from
the material causes it to lose buoyancy and sink due to its higher mean
molecular weight. The mixing continues until the molecular weight inversion is removed.
Thermohaline mixing is important in the collision products during the
contraction phase in the cases where material from the core of the primary
overlies material from the secondary, which happens for cases \emph{P} and
\emph{S}, discussed in \S \ref{sec:composition_profiles} and shown in
Figure \ref{fig:kippenhahn}.
By the time the collision products reach the main sequence 
the molecular weight inversion has been smoothed out. As a result more 
helium has been mixed into the centre of the collision product than would
have been the case if thermohaline mixing had been ignored. This reduces
the amount of hydrogen available for nuclear burning and decreases the
lifetime of the collision product.

A hydrogen burning shell at the edge of the core can drive a convection
zone that mixes in material from the layers above. This material can be
either hydrogen-rich or helium rich, depending on the particular case.
The convection zone can connect to the thermohaline layer, which means extra helium is mixed into the inner region of the star.

When the convective core is formed it can in turn connect to 
the convection zone driven by the burning shell (Figure
\ref{fig:kippenhahn}, lower left panel). If there is no burning shell,
it can connect to the thermohaline layer (Figure \ref{fig:kippenhahn} upper
left panel). In either case the end result is that
the interior of the collision product is mixed below the molecular weight
inversion. This produces a large central region that has been enhanced in helium
and is larger than the convective core itself. The net effect is comparable to that
of a star in which the convective core was larger initially but has shrunk.

Outside the mixed region, the composition profile is shallower than in a normal
star of the same mass.
This affects the structure of the star most strongly during the main
sequence although it continues to affect the star during later evolutionary
stages.


\subsection{Main sequence evolution}
\begin{figure}
\includegraphics[width=0.5\textwidth]{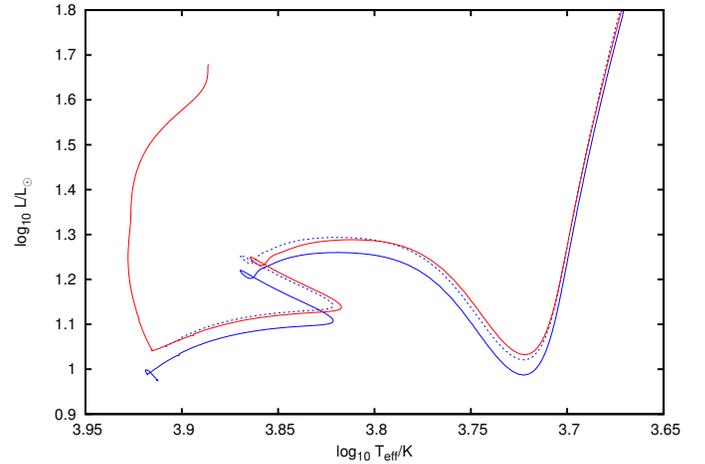}
\caption{Evolution track of the $t=2800, q=0.8, M=1.9$ remnant with a mass
of $1.75 M_\odot$ (top solid line) compared with the evolution track of a
normal star of the same mass (bottom solid line). The shape of the evolution
tracks after the contraction of the merger remnant is very similar, but the
merger track is offset to higher luminosity and is slightly cooler at the
end of the main sequence. The dashed line is a shifted version of the main
sequence track according to equations (\ref{eqn:luminosity_homology}) and
(\ref{eqn:mu_ratio_parameters}), see \S \ref{sec:recipe}.}
\label{fig:hrd_shifted}
\label{fig:hrd_shifted_analytic}
\end{figure}

Once the merger remnant has attained thermal equilibrium its further evolution proceeds
similarly to that of a normal star of the same mass. In particular, the
shape of the evolution track in a Hertzsprung-Russell diagram is the same.
The composition profile modifies the evolution mainly in two ways: by
affecting the lifetime and by affecting the luminosity. A typical evolution
track is shown in Figure \ref{fig:hrd_shifted}.

The luminosity is affected by the higher helium content of the envelope.
This affects the structure of the star through the equation of state (due
to the mean molecular weight) and the opacity, as discussed in Paper I, and results in a
larger radius and higher luminosity. Because the surface layers do not show
helium enhancement the effective temperature is not strongly affected,
although there is a systematic trend that the coolest part of the track
(just before hydrogen exhaustion at the end of the main sequence) is
shifted to slightly lower temperatures. Shifts in the luminosity
and effective temperature are listed in Table \ref{tab:model_table} for all
our merger remnants.

The remaining lifetime of the star is reduced by the higher
luminosity, but also by the central hydrogen abundance which is lower than
the abundance in the envelope. Because of this the merger remnants do not
resemble zero-age main sequence (ZAMS) stars after they settle down and are
both redder and brighter than ZAMS stars of the same mass.

While on the Hertzsprung gap tracks of the merger remnants are still
brighter than tracks of normal stars of the same mass. On the giant branch
the difference between the two tracks vanishes. When the first dredge-up
occurs the convective envelope enters the layer where the composition had
been modified by the merger. This enhances the post-dredge-up abundances of
helium and nitrogen and reduces the abundance of carbon compared to a
normal red giant (see Paper I for more details).


\subsection{The effect of lower metallicity}
The heavy-element abundance $Z$ affects the structure and evolution of
stars. In general stars with a lower $Z$ are hotter, more luminous and more
compact than stars of the same mass at higher $Z$. The mass at which a
convective core develops is higher and the lifetime is reduced.

This changes the structure of merger remnants in subtle ways.
\refereebf{For instance, the relative rate at which the progenitor stars evolve
along the main sequence is different for $Z=0.02$ and $Z=0.001$. This means
that if we consider a $Z=0.001$ collision at a time when the primary has
passed the same fraction of its main sequence lifetime, the secondary will
not have spent the same fraction of its main sequence lifetime as in the
$Z=0.02$ collision.}
Qualitatively \refereebf{the description given in the previous subsections}
remains valid for smaller $Z$. The data for the merger remnants calculated
for our grid are listed in Table \ref{tab:model_z001_table}.

The mass loss from the collisions is slightly lower than for the $Z=0.02$
grid although $C=0.3$ is still a reasonable guess. The reason for this can
be seen from (\ref{eqn:massloss}): the stars at lower $Z$ are more
centrally condensed than at higher $Z$ and the ratio $({R_{1,0.86} +
R_{2,0.86}})/({R_{1,0.5} + R_{2,0.5}})$ is smaller.

\section{HRD distributions: comparison with M67 and NGC188}
\label{sec:comparison}

In this section we compare the predicted locations of the collision
products from our grid with the observed blue stragglers in the open
clusters M67 and NGC188. For both of these clusters
\citet{article:pols_evmodels} determined isochrone fits. For consistency we
employ the same cluster parameters they used, as listed in Table~\ref{tab:cluster_parameters}.

\begin{table}
\caption{Ages determined from isochrone fitting for adopted values of
distance modulus and reddening. See \citet{article:pols_evmodels} for
details.}
\label{tab:cluster_parameters}
\begin{tabular}{llllll}
\hline
Name        & $\log_{10} \mathrm{age/yr}$ & $M_\mathrm{to}$ &  [Fe/H] & $m-M$   & $E(B-V)$\\
\hline\hline
M67         & $9.60$                      & $1.29$ & -0.06  & $ 9.60$ & $0.032$ \\
NGC 188     & $9.76$                      & $1.18$ & -0.06  & $11.40$ & $0.12$ \\ 
\hline
\end{tabular}
\end{table}

In order to make the comparison between our models and the observations it
is necessary to convert the theoretical surface parameters $(L,
T_\mathrm{eff}, \log g)$ to the observable magnitude and colour
$(M_\mathrm{V}, B-V)$. This conversion is done using the atmosphere models
from \citet{article:kurucz} with empirical corrections from
\citet{article:lejeune_calibration_of_spectra}.

\begin{figure}
\includegraphics[width=0.5\textwidth]{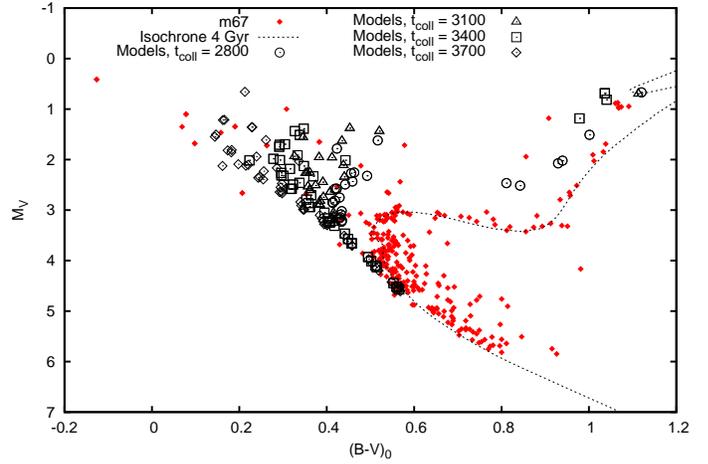}
\caption{Observed colour-magnitude diagram of the open cluster M67, with blue
stragglers highlighted and the positions of the stars in our collision
grid at the age of M67 overplotted. Different symbols correspond to
different times of collision, as indicated.}
\label{fig:m67_hrd}
\end{figure}

Figure \ref{fig:m67_hrd} shows the observed colour-magnitude diagram of
M67 ($\bullet$). The $V$ and $B-V$ data is taken from
\citet{article:cmd_m67} supplemented with \citet{montgomery_m67} for blue
stragglers that were missing from the \citet{article:cmd_m67} data.
Membership information (based on proper motions) is taken from
\citet{article:membership_m67}. \citet{article:cmd_m67} gives a distance
modulus $m-M = 9.60$ and reddening $E(B-V) = 0.03$, consistent with the
values used by \citet{article:pols_evmodels} to construct the $4
\,\mathrm{Gyr}$ isochrone. The location of the collision remnants at
$4 \,\mathrm{Gyr}$ are plotted with different symbols, corresponding to the
different times of collision. It is important to stress that
our model is not a population synthesis model, so one should not draw
conclusions from the density of the model points.

Our models lie in the observed blue straggler region. Some of the 
lower-mass merger remnants with masses below the turnoff appear among 
the normal main-sequence stars.  The bluest blue
straggler 
\citep[S977 \refereebf{with $M_\mathrm{V} = 0.413$ and $B-V = -0.126$},][]{mathys_bss_m67} is clearly outside the
region covered by our models, but it is thought to have a mass larger than
twice the turnoff mass \citep{sandquist_bss_starclusters} and
therefore cannot be formed by a single collision event. A few other
blue stragglers are somewhat bluer than the region covered by our models. 
They are apparently single stars and are thus very likely to be the
result of a stellar merger. Their blue positions can be explained if
they have undergone much stronger internal mixing than our models, or
if they have been formed more recently and are in that sense younger
than our models.  These blue stragglers are all slow rotators ($v\sin
i \leq 120$\,km/s) compared to normal stars of the same spectral type
\citep{mathys_bss_m67}, suggesting that rotational mixing has not
played a major role in their evolution. We therefore prefer the second
explanation, which suggests that M67 is dynamically active, in other
words, collisional blue stragglers are still being formed.  This
conclusion is consistent with the $N$-body model of
\citet{article:hurley_m67}.

A number of our merger remnants have evolved beyond the end of the
main sequence and appear on the giant branch but to the blue of the
normal cluster giant branch, indicating that merger remnants can still
stand out in the CMD even after the main sequence has ended. Two stars
in M67 are located in this region (S1040 and S1273); both are
spectroscopic binaries as well as unusual X-ray sources
\citep{vandenberg_1999A&A...347..866V}. It is not clear whether these
stars are merger remnants, their unusual location may also be the
result of a superposition of a giant and a turnoff star. In the case
of S1040 the companion is probably a white dwarf and the orbit is
circular \citep{mathieu_1990AJ....100.1859M}, so the system has likely
undergone mass transfer.

\begin{figure}
\includegraphics[width=0.5\textwidth]{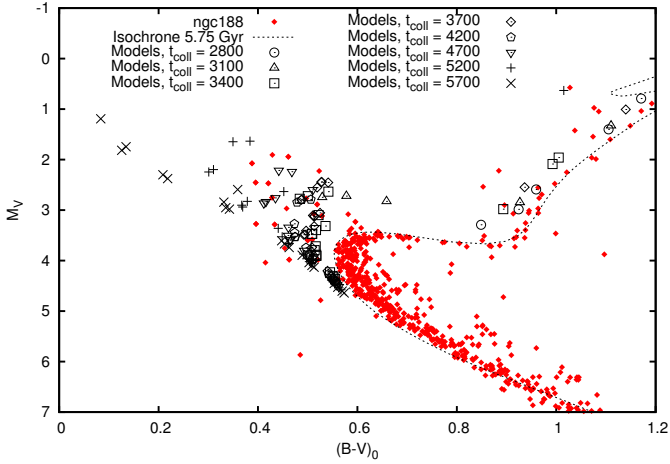}
\caption{As Figure \ref{fig:m67_hrd} for NGC 188}\label{fig:ngc188_hrd}
\end{figure}

Figure \ref{fig:ngc188_hrd} shows the comparison with NGC188, with $V$, 
$B-V$ and membership probabilities taken from
\citet{article:observations_ngc188}. Although NGC188 is known to be older
than M67, different authors give different age
estimates\citep{photometry_ngc188,article:observations_ngc188,five_open_clusters}.
We adopt $m-M = 11.4$ and $E(B-V) = 0.12$ and an age of $5.8
\,\mathrm{Gyr}$, as in \citet{article:pols_evmodels}. By contrast,
\citet{bonatto_ngc188} find $m-M = 11.1$ and $E(B-V) = 0.0$ from
fitting a $t=7 \,\mathrm{Gyr}$ Padova isochrone. 

Our models fill a larger region than the observed blue
straggler region, indicating that NGC188 has not formed massive blue
stragglers recently. This suggests that it is less dynamically active than
M67.
As for M67 we note that a number of merger remnants are beyond the main
sequence and on the giant branch, again parallel to the cluster giant
branch.

\section{Analytical description of results}
\label{sec:recipe}








To quantify the results of our collision calculations and compare the evolution
tracks of the collision products with evolution tracks of normal single stars 
it is necessary to specify which models should be compared.
In particular, we need to define an equivalent main sequence star
model.

Apart from the mass $M$ of the collision product the remaining lifetime is
a good parameter to consider. For this reason it is convenient to introduce
the fractional age $f$, which is the age of a star expressed in units of 
its main sequence lifetime,
\beq
f = \frac{t}{\tau_\mathrm{MS}}
\eeq
where $t$ is the absolute age and $\tau_\mathrm{MS} = \tau_\mathrm{MS}(M, Z)$
is the main sequence lifetime of a star of mass $M$ and composition specified 
by $Z$. As long as the star is on the main sequence, $0<f<1$.
We define the apparent age $f_\mathrm{app}$ of the collision product as the fractional
age of a normal main sequence star with the same remaining lifetime
$t_\mathrm{MS}$. These two quantities are related by
\beq
t_\mathrm{MS} = \tau_\mathrm{MS} (1-f_\mathrm{app}).
\eeq
In practice, we know $t_\mathrm{MS}$ from the detailed models (see Table
\ref{tab:model_table}) for the collision products and we can thus determine
$f_\mathrm{app}$. If we can predict $f_\mathrm{app}$ from the global
properties of the parent stars we can predict the collision product
lifetime. In the following we will describe a formalism that allows us to
do this using only standard stellar models and we give a set of
transformation rules that can be used to transform a standard stellar
evolution track to approximate the track of a merger remnant. By a standard
evolution track we mean the evolution track of a single star that is
evolved from the zero-age main sequence. For the standard tracks one can
use the \citet{article:hurley_sse} recipe, or interpolate on a grid of
single star models.
Our recipe can be used to improve the accuracy of evolution tracks for
merger remnants in star cluster simulations.

\subsection{Collision product lifetimes}
In normal main sequence stars there is a unique relation between the amount
of hydrogen and the age of the star. Collision products show a similar 
relation%
. Although 
it might seem better to relate the age of the star to the amount of hydrogen
in the core rather than the total amount of hydrogen, in practice this is not
a good measure of the age of the star for collision products because mixing
as described in \S \ref{sec:mixing} can increase the central hydrogen abundance
compared to its post-collision value.

\begin{figure}
   \center \includegraphics[angle=\imgangle,width=0.5\textwidth]{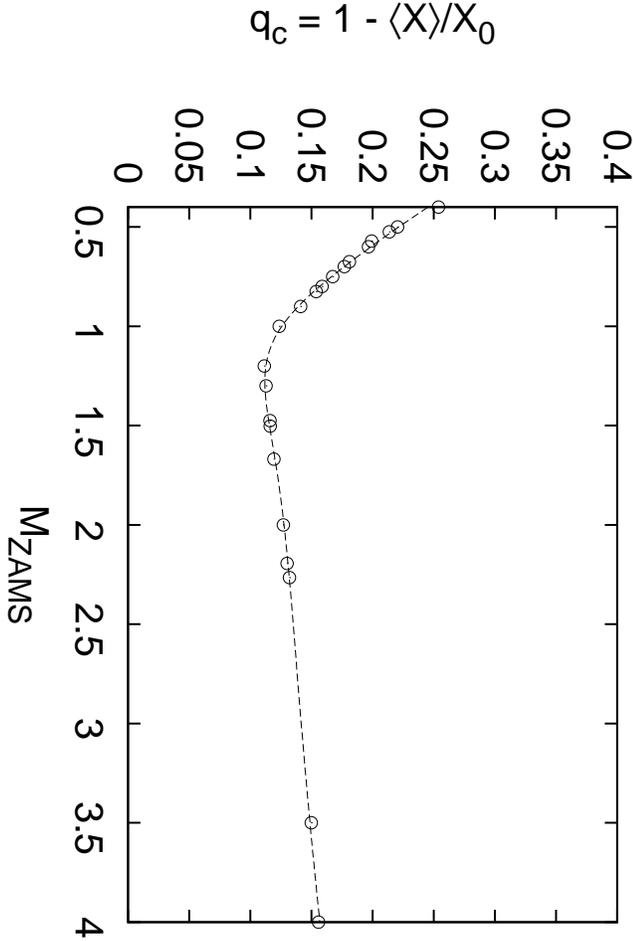}
   \caption{Effective core mass fraction for main sequence stars
   against ZAMS mass, for $Z=0.02$ ($\odot$). The fitting
   formula~(\ref{eq:qc_fit}) in the mass range $0.4$ -- $4.0$
   $M_\odot$ is shown as a dashed line.}
   \label{fig:qc_m}
\end{figure}

The relation between the amount of hydrogen and the age of the star is the
starting point for a relation between the global properties of the parent stars
and the apparent age $f_\mathrm{app}$ of the collisions product.
Let's consider a simple stellar model that divides the star in two parts: a 
core where hydrogen is burned to helium and an envelope which is not affected
by nuclear burning on the main sequence. The core has a mass fraction 
$q_\mathrm{c}$ of the total mass $M$. Such a model is not really applicable to 
low-mass stars, which do not have a well-defined core.
However, we can generalise the model by defining $q_\mathrm{c}$ to be the 
fraction of hydrogen that is burned during the main sequence, so that
\beq
q_\mathrm{c} = \frac{X_0 - \avg{X}_\mathrm{TAMS}}{X_0},
\eeq
with $X_0$ the initial (ZAMS) hydrogen fraction. We can think of $q_\mathrm{c}$
as being the \emph{effective core} mass fraction. If we approximate the nuclear
burning rate as steady, the average (total) amount of hydrogen follows from
the relative age $f$ as
\beq\label{eqn:hydrogen_against_f}
\avg{X}(t) = X_0 \left( 1 - q_\mathrm{c} f\right).
\eeq
In practice, $q_\mathrm{c}$ will depend on the stellar mass $M$ and
composition $Z$ and needs to be found from detailed stellar models. It is convenient to
have an analytic approximation for $q_\mathrm{c}(M)$. We use the
rational function
\beq \label{eq:qc_fit}
q_\mathrm{c}(M) = \frac{
   1 + c_1 M + c_2 M^3 + c_3 M^5 + c_4 M^7
}{
   c_5 + c_6 M^2 + c_7 M^4 + c_8 M^6
}
\qquad \mbox{($M$ in $M_\odot$)}.
\eeq
The coefficients $c_i$ have been found by least-squares fitting to
detailed models calculated with the STARS code for masses in the range
$M=0.4 \ldots 75\, M_\odot$. These models were computed without
convective overshooting. The fitting coefficients are
listed in Table~\ref{tab:qc}. In the mass range $0.4 \ldots 4.0\,
M_\odot$ a somewhat more accurate fit is possible without the terms
involving $M^6$ and $M^7$ (second and fourth column in
Table~\ref{tab:qc}); the quality of this fit for $Z=0.02$ can be
checked against Figure~\ref{fig:qc_m}. It is of course also possible
to interpolate on a table of detailed models.

\begin{table}
   \caption{Coefficients for the fitting formula~(\ref{eq:qc_fit}) for 
   $q_\mathrm{c}$. The first column lists values that are valid in the
   mass range $0.4$ -- $75$ $M_\odot$ for $Z=0.02$, the second column
   lists coefficients that give a more accurate fit in the mass range
   $0.4$ -- $4.0$ $M_\odot$. The third and fourth columns are similar
   but for $Z=0.001$. The bottom row (rms) gives the root mean square
   error of the fits.}
   \label{tab:qc}
   \center
   \begin{tabular}{lllll}
   \hline \hline
       & $Z=0.02$ & & $Z=0.001$ & \\
   $c$ & $0.4-75\,M_\odot$ & $0.4-4\,M_\odot$
       & $0.4-75\,M_\odot$ & $0.4-5\,M_\odot$  \\
   \hline
   $c_1$ & $-0.685213$    & $-0.999710$  &  $-0.682671$    & $-0.873596$\\
   $c_2$ & $ 0.289269$    & $ 0.276843$  &  $ 0.185036$    & $ 0.165455$\\
   $c_3$ & $ 0.0123223$   & $ 0.0393513$ &  $ 0.00658021$  & $ 0.0158297$\\
   $c_4$ & $ 3.3357$e$-6$ & $ 0$         &  $ 7.5450$e$-7$ & $ 0$\\
   $c_5$ & $ 2.70964$     & $ 2.75579$   &  $ 2.71437$     & $ 2.72675$\\
   $c_6$ & $ 1.44963$     & $-1.70434$   &  $ 0.334367$    & $-1.33733$\\
   $c_7$ & $ 0.629121$    & $ 1.46625$   &  $ 0.343889$    & $ 0.644498$\\
   $c_8$ & $ 0.000493117$ & $ 0$         &  $ 0.000185160$ & $ 0$\\
   \hline
   rms   & 1.4\%          & 1.2\%        &  1.7\%          & 1.3\% \\
   \hline
   \end{tabular}
\end{table}

To find an expression for $f_\mathrm{app}$ we first need to know the amount of
hydrogen in the collision product. To do this correctly we should consider the 
material that is ejected during the collision. This comes from the progenitor
star envelopes and therefore has the initial ZAMS hydrogen abundance $X_0$.
If $\phi$ is the fraction of mass lost during the collision (see \S
\ref{sec:massloss}) then the mass $M$ of the collision product is
$M=(1-\phi)(M_1+M_2)$. The average hydrogen abundance $\avg{X}$ in the
collision product immediately after the collision is then given by
\beq
M \avg{X} = M_1 \avg{X}_1(t) + M_2 \avg{X}_2(t) - \phi(M_1 + M_2) X_0.
\eeq
Inserting (\ref{eqn:hydrogen_against_f}) for the hydrogen abundance in the
parent stars and solving for $\avg{X}$,
\beq\label{eqn:hydrogen_merger}
\avg{X} = X_0 \left (
   1 - \frac{1}{1-\phi} 
   \frac{
      q_{\mathrm{c},1}f_1 + q_{\mathrm{c},2}f_2 q
   }{
      1+q
   } 
   \right).
\eeq
The equivalent age $f_\mathrm{app}$ is determined by the amount of hydrogen
$q_\mathrm{M}$ that gets mixed into the burning region through
\beq\label{eqn:hydrogen_merger_fapp}
\avg{X} = X_0 (1-q_\mathrm{M} f_\mathrm{app}).
\eeq
As already mentioned in \S \ref{sec:mixing} the larger fraction of processed
material within the merger remnant has an effect that is comparable to that
of a star with an initially larger core mass.
Setting $q_\mathrm{M} = \alpha q_\mathrm{c}(M)$ , we find $f_\mathrm{app}$ from
(\ref{eqn:hydrogen_merger}) as
\beq\label{eqn:fapp_pred}
f_\mathrm{app}  = 
\frac{1}{\alpha q_\mathrm{c}(M)}
\frac{1}{1-\phi}
\frac{ q_{\mathrm{c},1} f_1 + q_{\mathrm{c},2} f_1 q} { 1+q }.
\eeq
The parameter $\alpha$ parametrises the amount of mixing during the collision
and settling to the main sequence, with $\alpha = 1$ meaning no mixing at all
and $\alpha = 1/q_\mathrm{c}(M)$ meaning homogeneous mixing.
Our expression (\ref{eqn:fapp_pred}) is a generalisation of equation (69) in
\citet{article:tout_evolution_models} and (80) in \citet{article:hurley_bse},
which are recovered in the special case $\phi=0$,
$q_{\mathrm{c},1}=q_{\mathrm{c},2}=q_{\mathrm{c}}(M) = 1/10$ and 
$\alpha = 1/q_\mathrm{c}(M)$.
\begin{figure}
\includegraphics[angle=0,width=0.5\textwidth]{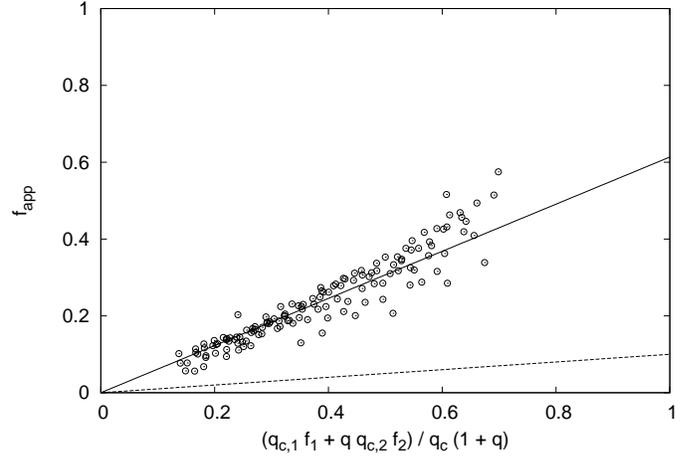}
\caption{The predicted apparent age according to (\ref{eqn:fapp_pred}),
for $\alpha=1.67$ (solid line). The detailed models are indicated by
$\odot$. The dashed line is the prescription from \citet{article:hurley_sse}.
}\label{fig:lifetime_prediction}
\end{figure}

The parameter $\alpha$ can be determined from the models in our grid because
all other factors in equation (\ref{eqn:fapp_pred}) are known. In principle
it should depend on the evolutionary stages of the two stars involved in the
collision, but in practice the value $\alpha=1.67$ works well for all models
in our $Z=0.02$ grid, while $\alpha = 1.43$ works well for our $Z=0.001$
grid.

Figure \ref{fig:lifetime_prediction} compares the predicted lifetime according
to (\ref{eqn:fapp_pred}) to the lifetime from our detailed models. Overall
agreement is very good, although the detailed models show more scatter.
The \refereebf{dashed} line is the prescription from \citet{article:hurley_sse}, which
predicts much longer lifetimes, especially for collisions involving very evolved
stars. This increases the predicted number of observable blue stragglers
formed through collisions by up to a factor $2$ compared to our detailed
models.

\subsection{Collision product luminosities}
The luminosity of the collision products follows very well from the homology
relation \citep{book:kippenhahn_weigert}
\beq\label{eqn:luminosity_homology}
L_\mathrm{merger} = L_\mathrm{MS} \left(\frac{\mu_\mathrm{merger}}{\mu_\mathrm{MS}}\right)^4.
\eeq
Here, $\mu_\mathrm{merger}$ is the average mean molecular weight in the collision
product at the start of the main sequence, which is the same as the average
mean molecular weight after the collision because the contraction phase is
fast enough that composition changes due to nuclear reactions can be
ignored. The average mean molecular weight of the main sequence model
$\mu_\mathrm{MS}$ needs to be chosen at the equivalent stage of evolution.
Because the shift in effective temperature is small we can choose the point
in the evolution track with the same $T_\mathrm{eff}$ as the equivalent
point. The averages of the mean molecular weight need to be calculated with
respect to the mass (as opposed to the radius), which is the same as
averaging over all particles.

It is possible to get an estimate for the mean molecular weight using the simple
model (\ref{eqn:hydrogen_against_f}). For a fully ionised hydrogen/helium 
mixture, the mean molecular weight can be approximated by (\emph{e.g.}
\citet{book:cox_and_giuli})
\beq
\mu^{-1} = 2 X + \frac{3}{4} Y + \frac{1}{2} Z,
\eeq
or, by eliminating the helium abundance $Y$,
\beq
\mu^{-1} = \frac{1}{4} \left(
5 X + 3 - Z
\right).
\eeq
Then
\beq\label{eqn:mu_ratio}
\frac{
   {\mu}_\mathrm{merger}
}{
   {\mu}_\mathrm{MS}
}
=
\frac{
   5 \avg{X}_\mathrm{MS} + 3 - Z
}{
   5 \avg{X}_\mathrm{merger} + 3 - Z
}.
\eeq
For the main sequence star we can use (\ref{eqn:hydrogen_against_f}) directly
but to match the hydrogen abundance of the merger remnant we use a modified 
version of (\ref{eqn:hydrogen_against_f}),
\beq\label{eqn:match_hydrogen_against_f}
\avg{X}_\mathrm{merger} = X_0 \left( 1 - \beta - \alpha q_\mathrm{c} f_\mathrm{app} \right).
\eeq
Physically, this extra parameter is necessary because $f_\mathrm{app}$ was
calibrated to the main sequence timescale $\tau_\mathrm{MS}$ for a normal
star. The collision products behave more like stars with a different initial
composition and can have a different main sequence timescale. In other words,
$f_\mathrm{app}$ does not represent the evolutionary stage of the collision
product, just its remaining lifetime.
Mathematically, the need for an extra parameter arises because we are now
fitting two quantities (the lifetime and the hydrogen abundance).

The offset should vanish if the progenitor stars were ZAMS stars 
(in which case $f_\mathrm{app} = 0$) or if there was no mixing during the
collision ($\alpha = 1$). This is satisfied if 
$\beta \propto (\alpha-1) f_\mathrm{app}$. We find that setting
$\beta = (\alpha-1) q_\mathrm{c} f_\mathrm{app} / (1-q_\mathrm{c})$ gives
a good match for all collisions in our grid.
Inserting this expression in (\ref{eqn:match_hydrogen_against_f}) gives
\beq
\avg{X}_\mathrm{merger} = X_0 \left( 1 - \alpha q_\mathrm{c} \frac{\alpha (2-q_\mathrm{c})-1}{\alpha (1-q_\mathrm{c})} f_\mathrm{app} \right).
\eeq
Setting $\delta \avg{X} = \avg{X}_\mathrm{merger}-\avg{X}_\mathrm{MS}$
and introducing $\mu_0 = 5 \avg{X}_\mathrm{MS} + 3 - Z$,
\beq\label{eqn:mu_ratio_parameters}
\frac{
   {\mu}_\mathrm{merger}
}{
   {\mu}_\mathrm{MS}
}
=\frac{\mu_0}{\mu_0 + 5\delta\avg{X}}
\approx 1-5\frac{\delta\avg{X}}{\mu_0}.
\eeq
From (\ref{eqn:hydrogen_against_f}) and (\ref{eqn:hydrogen_merger_fapp}),
\beq
\delta\avg{X} 
=
X_0 q_c f_{app}
\frac{(2-q_\mathrm{c}) (1-\alpha)}{1-q_\mathrm{c}}.
\eeq
Physically, $\alpha>1$ and we see that in that case 
$\avg{\mu}_\mathrm{merger}>\avg{\mu}_\mathrm{MS}$, as we would expect.

The luminosity shift predicted by our analytic prescription is shown in
Figure \ref{fig:hrd_shifted_analytic} (dashed line).
The main sequence part of the track is very well represented by our
prescription, although the hook at the end is slightly too hot. Even the
Hertzsprung gap and the giant branch are very well matched with the same
shift.

\subsection{Collision product effective temperatures and radii}
The effective temperature of the merger remnants is not very different
from that in a normal star of the same mass (see the listed temperature
shifts in Table \ref{tab:model_table}), but the reddest point of the
main sequence hook is slightly cooler.

Since the effective temperature of the collision products is very similar
to that of normal stars it is possible to use the effective temperature
expected for normal stars. Taking the luminosity from (\ref{eqn:luminosity_homology})
the radius of the collision product follows from the Stefan-Boltzmann law,
so that
\beq
R_\mathrm{merger} = R_\mathrm{MS}
\left(\frac{\mu_\mathrm{merger}}{\mu_\mathrm{MS}}\right)^2.
\eeq

\section{Discussion and conclusions}
\label{sec:discussion}

We have calculated a grid of stellar evolution models for the remnants of
collisions between main sequence stars. These models improve on the models
in the literature by using a better model for the initial structure of the
merger remnant \citep{article:bailyn_pinsonneault} or by studying a larger
portion of the $t, q, M$ parameter space \citep{article:sills_evcolprod1}. 

We have compared the position of our models in the Hertzsprung-Russell
diagram with the observed blue straggler regions in the open clusters
M67 and NGC 188 and found that our models can populate the blue straggler
region, provided that the blue stragglers in M67 are relatively `young' and
those in NGC 188 are relatively `old'.

We have used our results to formulate a recipe that can be used to
transform a normal evolution track, which can be obtained from fitting
formulae \citep{article:hurley_sse} or by interpolation on a table
\citep[\emph{e.g.}][]{article:pols_evmodels,padova_tracks_2002}. This can
be used to improve the treatment of collisions in $N$-body calculations and
population modelling.

Our models do not include the effect of rotation. Rotation can modify our
results in two ways: collision products need to lose angular momentum and
therefore mass to reach hydrostatic equilibrium and rapid rotation can
induce mixing.
It is clear that for a collision product to survive angular momentum must
be lost but the mechanism by which this occurs is unclear.
Magnetic fields may be involved, but their role is not well understood.
Rotational mixing can enrich the stellar envelope in helium, increasing
the luminosity and extending the stellar lifetime. By ignoring rotation we
have assumed that an efficient mechanism to remove the angular momentum
operates that spins the star down sufficiently that rotational mixing is
not important for the long-term evolution. The fact that blue stragglers in
M67 are observed to be rotating slowly \citep{mathys_bss_m67} can be
interpreted as an indication that such an efficient mechanism for removing
angular momentum does operate in reality.

Although our models can fill the observed blue straggler region in M67 and
NGC 188, this does not directly give information about the expected
distribution of blue stragglers in the colour magnitude diagram because the
models need to be weighted with the probability for each collision. This
requires a dynamical (or statistical, \emph{e.g.} Monte-Carlo) model for
the cluster evolution.

There are several directions for future research that could be explored.
From the side of cluster modelling it is interesting to study the effect of
our predicted corrections to the evolution tracks on the number of
expected blue stragglers and on the blue straggler luminosity function.
On the topic of detailed modelling of merger remnant evolution,
including rotation and rotational mixing \citep[extending the work of][]{
sills_evcolprod2,sills_evcolprod3} is a logical next
step. However, this needs to be coupled with a mechanism for angular
momentum loss from the merger remnant, which requires a better
understanding of mass loss from rapidly rotating low mass
stars, possibly involving magnetic fields as well. A simple equilibrium
dynamo may not be adequate to model the magnetic field of a merger remnant,
which might be very different from the magnetic field of a normal star
following the collision. Magnetohydrodynamic models of stellar mergers and
stellar evolution models including magnetic fields are also relevant in
this light and are a logical next step.


%
\begin{acknowledgements}
We thank \refereebf{the referee, Alison Sills, for valuable comments and}
Frank Verbunt for useful discussion on the comparison with observations of
M67.  EG is supported by NWO under grant 614.000.303
\end{acknowledgements}

\bibliographystyle{aa}
\bibliography{low_mass_grid}

\Online
\addtocounter{table}{1}
\onltab{\arabic{table}}{
\longtab{\arabic{table}}{

\begin{longtable}{rrrrrrrrrrrrrr}
\caption{Results for the $Z=0.02$ collisions. The first three columns list the time
of collision $t$ (in Myr), the mass ratio $q$ and the total initial mass
$M$ (in $M_\odot$). Column four gives the case for the collision (M, S or
P, see \S \ref{sec:composition_profiles}). The total remnant mass
$M_\mathrm{\remnantabbr}$ is smaller than the initial mass $M$ because of mass
loss in the collision. Column six gives the constant in the expression for
the mass loss (\ref{eqn:massloss_approx}) for each of the individual
collisions. Columns seven, eight and nine give the main sequence lifetime
of the primary ($\tau_\mathrm{ms,1}$), secondary ($\tau_\mathrm{ms,2}$) and
of a star of mass $M_\mathrm{\remnantabbr}$ ($\tau_\mathrm{ms,2}$). Column ten
gives the actual remaining main sequence lifetime $t_\mathrm{ms}$ of the collision
product. Columns eleven and twelve give the central abundance of hydrogen
after the collision ($X_\mathrm{c,0}$) and at the beginning of the main
sequence phase ($X_\mathrm{c,zms}$). The final two columns give the change in
$\log_{10}$ luminosity $L$ (in solar units) and $\log_{10}$ effective
temperature $T_\mathrm{eff}$ (in $\mathrm{K}$) at the reddest point of the
main sequence track (shortly before core hydrogen exhaustion).
}\label{tab:model_table}\\
\hline\hline
$t$ & $q$ & $M$ & Case & $M_\mathrm{\remnantabbr}$ & $\frac{(1+q)^2}{q}\phi$ &
$\tau_\mathrm{ms, 1}$ & $\tau_\mathrm{ms, 2}$ & $\tau_\mathrm{ms}$ &
$t_\mathrm{ms}$ & $X_{\mathrm{c},0}$ & $X_{\mathrm{c,zms}}$ & $\Delta
\logten L$ & $\Delta \logten T_\mathrm{eff}$\\
\hline
\endfirsthead

\caption{cont.}\\
\hline\hline
$t$ & $q$ & $M$ & Case & $M_\mathrm{\remnantabbr}$ & $\frac{(1+q)^2}{q}\phi$ &
$\tau_\mathrm{ms, 1}$ & $\tau_\mathrm{ms, 2}$ & $\tau_\mathrm{ms}$ &
$t_\mathrm{ms}$ & $X_{\mathrm{c},0}$ & $X_{\mathrm{c,zms}}$ & $\Delta
\logten L$ & $\Delta \logten T_\mathrm{eff}$\\
\hline
\endhead

\hline
\endfoot
2800 & 0.4 & 1.5 & S & 1.41 & 0.28 &    6247 &  181307 &    2413 &    2100 & 0.691 & 0.676 &  0.053 & -0.004\\
2800 & 0.4 & 1.7 & P & 1.60 & 0.29 &    3337 &  134060 &    1726 &    1234 & 0.379 & 0.379 &  0.074 & -0.008\\
2800 & 0.4 & 1.8 & S & 1.69 & 0.31 &    3444 &  114628 &    1478 &     951 & 0.547 & 0.574 &  0.095 & -0.009\\
2800 & 0.4 & 1.9 & P & 1.77 & 0.32 &    2751 &   97623 &    1273 &     647 & 0.063 & 0.505 &  0.104 & -0.014\\
2800 & 0.5 & 1.5 & S & 1.41 & 0.27 &    8644 &  124021 &    2519 &    2180 & 0.689 & 0.672 &  0.028 & -0.002\\
2800 & 0.5 & 1.6 & S & 1.50 & 0.29 &    6382 &  102516 &    2103 &    1709 & 0.686 & 0.666 &  0.039 & -0.001\\
2800 & 0.5 & 1.7 & M & 1.59 & 0.29 &    4458 &   85148 &    1746 &    1395 & 0.490 & 0.490 &  0.050 & -0.006\\
2800 & 0.5 & 1.8 & P & 1.69 & 0.28 &    4206 &   70071 &    1474 &    1074 & 0.415 & 0.415 &  0.058 & -0.006\\
2800 & 0.5 & 1.9 & S & 1.78 & 0.28 &    3438 &   57523 &    1261 &    1126 & 0.677 & 0.671 &  0.028 & -0.002\\
2800 & 0.5 & 2.0 & P & 1.86 & 0.31 &    3001 &   47262 &    1104 &     534 & 0.124 & 0.128 &  0.065 & -0.003\\
2800 & 0.6 & 1.5 & S & 1.40 & 0.28 &   11552 &   87227 &    2608 &    2253 & 0.685 & 0.674 &  0.016 & -0.001\\
2800 & 0.6 & 1.6 & S & 1.49 & 0.31 &    8644 &   70071 &    2101 &    1848 & 0.683 & 0.662 &  0.032 & -0.001\\
2800 & 0.6 & 1.7 & S & 1.58 & 0.30 &    6504 &   56129 &    1782 &    1475 & 0.679 & 0.645 &  0.034 & -0.005\\
2800 & 0.6 & 1.8 & P & 1.68 & 0.28 &    4907 &   45016 &    1485 &    1152 & 0.513 & 0.513 &  0.041 & -0.005\\
2800 & 0.6 & 1.9 & P & 1.78 & 0.27 &    3734 &   36312 &    1266 &     906 & 0.422 & 0.422 &  0.049 & -0.005\\
2800 & 0.6 & 2.0 & P & 1.87 & 0.29 &    3461 &   29518 &    1099 &     711 & 0.349 & 0.348 &  0.060 & -0.005\\
2800 & 0.6 & 2.1 & P & 1.94 & 0.32 &    3159 &   24457 &     984 &       3 & 0.328 & 0.328 &  0.075 & -0.004\\
2800 & 0.6 & 2.2 & P & 2.04 & 0.30 &    2722 &   19931 &     859 &     431 & 0.048 & 0.048 &  0.084 & -0.008\\
2800 & 0.8 & 1.5 & S & 1.40 & 0.26 &   19036 &   47262 &    2493 &    2352 & 0.676 & 0.669 &  0.016 & -0.002\\
2800 & 0.8 & 1.6 & S & 1.49 & 0.29 &   14553 &   36598 &    2162 &    1910 & 0.672 & 0.664 &  0.014 & -0.001\\
2800 & 0.8 & 1.7 & S & 1.57 & 0.31 &   11182 &   28644 &    1836 &    1577 & 0.662 & 0.649 &  0.016 & -0.003\\
2800 & 0.8 & 1.8 & M & 1.67 & 0.29 &    8644 &   22659 &    1511 &    1259 & 0.586 & 0.586 &  0.021 & -0.004\\
2800 & 0.8 & 1.9 & P & 1.75 & 0.31 &    6712 &   18089 &    1315 &    1051 & 0.550 & 0.550 &  0.029 & -0.005\\
2800 & 0.8 & 2.0 & P & 1.86 & 0.29 &    5225 &   14553 &    1113 &     819 & 0.509 & 0.518 &  0.030 & -0.003\\
2800 & 0.8 & 2.1 & P & 1.94 & 0.31 &    4861 &   11784 &     986 &     707 & 0.456 & 0.532 &  0.051 & -0.007\\
2800 & 0.8 & 2.2 & P & 2.04 & 0.29 &    3956 &    9581 &     859 &     570 & 0.391 & 0.515 &  0.057 & -0.007\\
2800 & 0.8 & 2.3 & P & 2.11 & 0.33 &    3445 &    7811 &     777 &     470 & 0.322 & 0.477 &  0.071 & -0.008\\
2800 & 0.8 & 2.4 & P & 2.23 & 0.29 &    3001 &    6384 &     671 &     361 & 0.223 & 0.450 &  0.080 & -0.009\\
2800 & 0.9 & 1.5 & S & 1.39 & 0.30 &   23930 &   36719 &    2643 &    2439 & 0.668 & 0.669 &  0.009 & -0.001\\
2800 & 0.9 & 1.6 & S & 1.49 & 0.28 &   18301 &   28286 &    2135 &    1920 & 0.657 & 0.661 &  0.010 & -0.002\\
2800 & 0.9 & 1.7 & M & 1.57 & 0.31 &   14150 &   22050 &    1818 &    1586 & 0.649 & 0.649 &  0.013 & -0.002\\
2800 & 0.9 & 1.8 & M & 1.75 & 0.11 &   11029 &   17369 &    1324 &    1055 & 0.612 & 0.611 &  0.015 & -0.002\\
2800 & 0.9 & 1.9 & M & 1.76 & 0.31 &    8644 &   13797 &    1312 &    1071 & 0.585 & 0.620 &  0.024 & -0.005\\
2800 & 0.9 & 2.0 & P & 1.86 & 0.28 &    6801 &   11029 &    1111 &     859 & 0.554 & 0.539 &  0.026 & -0.005\\
2800 & 1.0 & 1.5 & M & 1.39 & 0.30 &   29518 &   29518 &    2689 &    2415 & 0.668 & 0.668 &  0.013 & -0.002\\
2800 & 1.0 & 1.6 & M & 1.49 & 0.28 &   22659 &   22659 &    2149 &    1924 & 0.652 & 0.659 &  0.008 & -0.002\\
2800 & 1.0 & 1.8 & M & 1.67 & 0.29 &   13797 &   13797 &    1512 &    1293 & 0.623 & 0.630 &  0.020 & -0.005\\
2800 & 1.0 & 1.9 & M & 1.75 & 0.31 &   10895 &   10895 &    1317 &    1093 & 0.605 & 0.613 &  0.028 & -0.005\\
2800 & 1.0 & 2.0 & M & 1.86 & 0.28 &    8644 &    8646 &    1110 &     854 & 0.580 & 0.591 &  0.026 & -0.004\\
2800 & 1.0 & 2.1 & M & 1.94 & 0.30 &    6885 &    6885 &     984 &     727 & 0.552 & 0.559 &  0.038 & -0.006\\
2800 & 1.0 & 2.2 & M & 2.05 & 0.27 &    5493 &    5493 &     848 &     583 & 0.514 & 0.518 &  0.046 & -0.006\\
2800 & 1.0 & 2.4 & M & 2.23 & 0.29 &    4206 &    4261 &     673 &     392 & 0.412 & 0.414 &  0.070 & -0.009\\
3100 & 0.4 & 1.5 & S & 1.41 & 0.29 &    6247 &  181307 &    2443 &    2064 & 0.682 & 0.669 &  0.058 & -0.004\\
3100 & 0.4 & 1.6 & M & 1.51 & 0.26 &    4537 &  156179 &    1869 &    1482 & 0.457 & 0.457 &  0.073 & -0.008\\
3100 & 0.4 & 1.7 & P & 1.60 & 0.30 &    3337 &  134060 &    1728 &    1143 & 0.324 & 0.324 &  0.079 & -0.008\\
3100 & 0.4 & 1.8 & P & 1.68 & 0.31 &    3444 &  114628 &    1487 &     864 & 0.204 & 0.204 &  0.102 & -0.008\\
3100 & 0.5 & 1.5 & S & 1.41 & 0.28 &    8644 &  124021 &    2538 &    2149 & 0.689 & 0.666 &  0.031 & -0.003\\
3100 & 0.5 & 1.6 & S & 1.50 & 0.27 &    6382 &  102516 &    2067 &    1674 & 0.659 & 0.645 &  0.041 & -0.006\\
3100 & 0.5 & 1.7 & P & 1.58 & 0.31 &    4458 &   85148 &    1771 &    1341 & 0.465 & 0.465 &  0.054 & -0.006\\
3100 & 0.5 & 1.8 & P & 1.69 & 0.28 &    4206 &   70071 &    1475 &    1017 & 0.365 & 0.365 &  0.064 & -0.005\\
3100 & 0.5 & 1.9 & S & 1.78 & 0.28 &    3438 &   57523 &    1261 &    1126 & 0.677 & 0.671 &  0.028 & -0.002\\
3100 & 0.5 & 2.0 & P & 1.86 & 0.31 &    3001 &   47262 &    1109 &     513 & 0.034 & 0.471 &  0.093 & -0.012\\
3100 & 0.6 & 1.5 & S & 1.40 & 0.29 &   11552 &   87227 &    2534 &    2295 & 0.684 & 0.677 &  0.033 & -0.001\\
3100 & 0.6 & 1.6 & S & 1.49 & 0.28 &    8644 &   70071 &    2057 &    1747 & 0.681 & 0.655 &  0.028 & -0.004\\
3100 & 0.6 & 1.7 & M & 1.58 & 0.31 &    6504 &   56129 &    1799 &    1447 & 0.540 & 0.540 &  0.036 & -0.006\\
3100 & 0.6 & 1.9 & P & 1.76 & 0.31 &    3734 &   36312 &    1301 &     885 & 0.387 & 0.387 &  0.056 & -0.007\\
3100 & 0.6 & 2.0 & P & 1.86 & 0.30 &    3461 &   29518 &    1108 &     673 & 0.286 & 0.286 &  0.068 & -0.004\\
3100 & 0.6 & 2.1 & P & 1.96 & 0.29 &    3159 &   24457 &     965 &     534 & 0.144 & 0.144 &  0.088 & -0.011\\
3100 & 0.8 & 1.5 & S & 1.40 & 0.27 &   19036 &   47262 &    2480 &    2340 & 0.673 & 0.666 &  0.016 & -0.002\\
3100 & 0.8 & 1.6 & S & 1.48 & 0.30 &   14553 &   36598 &    2116 &    1898 & 0.666 & 0.655 &  0.017 & -0.001\\
3100 & 0.8 & 1.7 & M & 1.59 & 0.27 &   11182 &   28644 &    1768 &    1511 & 0.666 & 0.642 &  0.019 & -0.005\\
3100 & 0.8 & 1.8 & M & 1.67 & 0.30 &    8644 &   22659 &    1523 &    1249 & 0.573 & 0.573 &  0.026 & -0.005\\
3100 & 0.8 & 1.9 & P & 1.77 & 0.27 &    6712 &   18089 &    1275 &     982 & 0.531 & 0.531 &  0.030 & -0.005\\
3100 & 0.8 & 2.0 & P & 1.86 & 0.28 &    5225 &   14553 &    1104 &     777 & 0.484 & 0.558 &  0.032 & -0.004\\
3100 & 0.8 & 2.1 & P & 1.93 & 0.32 &    4861 &   11784 &     997 &     680 & 0.422 & 0.508 &  0.057 & -0.007\\
3100 & 0.8 & 2.2 & P & 2.04 & 0.29 &    3956 &    9581 &     855 &     534 & 0.340 & 0.486 &  0.066 & -0.008\\
3100 & 0.8 & 2.3 & P & 2.14 & 0.29 &    3445 &    7811 &     756 &     434 & 0.248 & 0.463 &  0.079 & -0.008\\
3100 & 0.8 & 2.4 & P & 2.22 & 0.30 &    3001 &    6384 &     676 &       1 & 0.054 & 0.054 &  0.579 &  0.048\\
3100 & 0.9 & 1.7 & M & 1.57 & 0.31 &   14150 &   22050 &    1829 &    1568 & 0.646 & 0.643 &  0.014 & -0.002\\
3100 & 0.9 & 1.8 & M & 1.67 & 0.29 &   11029 &   17369 &    1515 &    1265 & 0.599 & 0.628 &  0.022 & -0.005\\
3100 & 0.9 & 2.0 & P & 1.85 & 0.29 &    6801 &   11029 &    1120 &     841 & 0.548 & 0.585 &  0.036 & -0.006\\
3100 & 1.0 & 1.5 & M & 1.38 & 0.31 &   29518 &   29518 &    2588 &    2388 & 0.664 & 0.662 &  0.015 &  0.001\\
3100 & 1.0 & 1.6 & M & 1.49 & 0.28 &   22659 &   22659 &    2095 &    1893 & 0.649 & 0.654 &  0.011 & -0.003\\
3100 & 1.0 & 1.7 & M & 1.57 & 0.31 &   18175 &   18175 &    1840 &    1575 & 0.633 & 0.642 &  0.014 & -0.003\\
3100 & 1.0 & 1.8 & M & 1.67 & 0.29 &   13797 &   13797 &    1520 &    1282 & 0.615 & 0.624 &  0.024 & -0.005\\
3100 & 1.0 & 1.9 & M & 1.77 & 0.27 &   10895 &   10895 &    1276 &    1038 & 0.593 & 0.601 &  0.031 & -0.006\\
3100 & 1.0 & 2.0 & M & 1.85 & 0.29 &    8644 &    8646 &    1122 &     847 & 0.566 & 0.586 &  0.033 & -0.006\\
3100 & 1.0 & 2.1 & M & 1.93 & 0.31 &    6885 &    6885 &     997 &     705 & 0.535 & 0.541 &  0.042 & -0.007\\
3100 & 1.0 & 2.2 & M & 2.04 & 0.29 &    5493 &    5493 &     862 &     566 & 0.491 & 0.488 &  0.053 & -0.007\\
3100 & 1.0 & 2.4 & M & 2.21 & 0.31 &    4206 &    4261 &     682 &     363 & 0.378 & 0.379 &  0.078 & -0.011\\
3400 & 0.4 & 1.2 & S & 1.13 & 0.29 &   16991 &  314052 &    5263 &    4416 & 0.698 & 0.690 &  0.023 &  0.000\\
3400 & 0.4 & 1.3 & S & 1.23 & 0.28 &   12092 &  243954 &    3952 &    3439 & 0.691 & 0.683 &  0.028 & -0.000\\
3400 & 0.4 & 1.4 & S & 1.32 & 0.27 &    8644 &  210585 &    3089 &    2568 & 0.690 & 0.676 &  0.042 & -0.002\\
3400 & 0.4 & 1.5 & S & 1.42 & 0.27 &    6247 &  181307 &    2495 &    1968 & 0.689 & 0.657 &  0.055 & -0.004\\
3400 & 0.4 & 1.6 & P & 1.50 & 0.30 &    4537 &  156179 &    2087 &    1486 & 0.411 & 0.411 &  0.070 & -0.006\\
3400 & 0.4 & 1.7 & P & 1.60 & 0.28 &    3337 &  134060 &    1710 &    1038 & 0.262 & 0.262 &  0.080 & -0.007\\
3400 & 0.4 & 1.8 & P & 1.69 & 0.29 &    3444 &  114628 &    1467 &     624 & 0.106 & 0.110 &  0.078 &  0.002\\
3400 & 0.5 & 1.5 & S & 1.40 & 0.29 &    8644 &  124021 &    2555 &    2128 & 0.681 & 0.661 &  0.034 & -0.003\\
3400 & 0.5 & 1.6 & M & 1.50 & 0.28 &    6382 &  102516 &    2088 &    1682 & 0.502 & 0.639 &  0.048 & -0.005\\
3400 & 0.5 & 1.7 & P & 1.59 & 0.30 &    4458 &   85148 &    1762 &    1268 & 0.430 & 0.430 &  0.058 & -0.007\\
3400 & 0.5 & 1.8 & P & 1.68 & 0.30 &    4206 &   70071 &    1486 &     927 & 0.309 & 0.310 &  0.069 & -0.006\\
3400 & 0.5 & 1.9 & S & 1.78 & 0.28 &    3438 &   57523 &    1262 &    1129 & 0.676 & 0.671 &  0.028 & -0.002\\
3400 & 0.6 & 1.2 & S & 1.12 & 0.27 &   29518 &  162623 &    5079 &    4986 & 0.689 & 0.688 &  0.022 & -0.001\\
3400 & 0.6 & 1.3 & S & 1.22 & 0.27 &   21253 &  133093 &    4005 &    3692 & 0.688 & 0.681 &  0.018 & -0.001\\
3400 & 0.6 & 1.4 & S & 1.31 & 0.26 &   15571 &  108190 &    3152 &    2812 & 0.686 & 0.674 &  0.018 & -0.001\\
3400 & 0.6 & 1.5 & S & 1.40 & 0.30 &   11552 &   87227 &    2539 &    2257 & 0.684 & 0.664 &  0.029 & -0.002\\
3400 & 0.6 & 1.6 & S & 1.49 & 0.29 &    8644 &   70071 &    2131 &    1721 & 0.679 & 0.648 &  0.030 & -0.004\\
3400 & 0.6 & 1.7 & P & 1.59 & 0.27 &    6504 &   56129 &    1746 &    1367 & 0.509 & 0.509 &  0.040 & -0.006\\
3400 & 0.6 & 1.8 & P & 1.68 & 0.28 &    4907 &   45016 &    1487 &    1065 & 0.440 & 0.440 &  0.049 & -0.006\\
3400 & 0.6 & 1.9 & P & 1.76 & 0.31 &    3734 &   36312 &    1301 &     830 & 0.335 & 0.335 &  0.060 & -0.007\\
3400 & 0.6 & 2.0 & P & 1.86 & 0.30 &    3461 &   29518 &    1115 &     606 & 0.206 & 0.203 &  0.071 & -0.004\\
3400 & 0.6 & 2.1 & P & 1.93 & 0.34 &    3159 &   24457 &    1000 &      12 & 0.037 & 0.037 &  0.095 & -0.009\\
3400 & 0.8 & 1.2 & S & 1.12 & 0.26 &   47280 &  102516 &    4972 &    3149 & 0.686 & 0.684 &  0.008 &  0.009\\
3400 & 0.8 & 1.3 & S & 1.21 & 0.29 &   34390 &   79923 &    4135 &      16 & 0.682 & 0.680 &  0.000 &  0.001\\
3400 & 0.8 & 1.4 & S & 1.30 & 0.29 &   25450 &   61494 &    3244 &    2995 & 0.679 & 0.673 &  0.011 & -0.001\\
3400 & 0.8 & 1.5 & S & 1.40 & 0.27 &   19036 &   47262 &    2492 &    2323 & 0.673 & 0.661 &  0.017 & -0.002\\
3400 & 0.8 & 1.6 & S & 1.48 & 0.30 &   14553 &   36598 &    2126 &    1886 & 0.667 & 0.651 &  0.018 & -0.002\\
3400 & 0.8 & 1.7 & M & 1.58 & 0.28 &   11182 &   28644 &    1780 &    1494 & 0.596 & 0.596 &  0.022 & -0.005\\
3400 & 0.8 & 1.8 & P & 1.68 & 0.28 &    8644 &   22659 &    1499 &    1192 & 0.553 & 0.553 &  0.028 & -0.005\\
3400 & 0.8 & 1.9 & P & 1.76 & 0.31 &    6712 &   18089 &    1311 &     968 & 0.513 & 0.510 &  0.031 & -0.005\\
3400 & 0.8 & 2.0 & P & 1.85 & 0.29 &    5225 &   14553 &    1120 &     782 & 0.455 & 0.559 &  0.046 & -0.007\\
3400 & 0.8 & 2.1 & P & 1.95 & 0.29 &    4861 &   11784 &     970 &     628 & 0.378 & 0.493 &  0.060 & -0.009\\
3400 & 0.8 & 2.2 & P & 2.03 & 0.32 &    3956 &    9581 &     877 &     503 & 0.297 & 0.452 &  0.075 & -0.009\\
3400 & 0.8 & 2.3 & P & 2.13 & 0.29 &    3445 &    7811 &     759 &     385 & 0.185 & 0.391 &  0.085 & -0.010\\
3400 & 0.8 & 2.4 & P & 2.18 & 0.37 &    3001 &    6384 &     709 &     287 & 0.201 & 0.201 &  0.109 & -0.015\\
3400 & 0.9 & 1.7 & M & 1.59 & 0.27 &   14150 &   22050 &    1764 &    1531 & 0.609 & 0.621 &  0.022 & -0.004\\
3400 & 0.9 & 1.8 & M & 1.67 & 0.30 &   11029 &   17369 &    1526 &    1245 & 0.586 & 0.584 &  0.023 & -0.006\\
3400 & 0.9 & 1.9 & P & 1.77 & 0.28 &    8644 &   13797 &    1280 &    1002 & 0.559 & 0.597 &  0.032 & -0.006\\
3400 & 0.9 & 2.0 & P & 1.86 & 0.28 &    6801 &   11029 &    1110 &     801 & 0.510 & 0.569 &  0.037 & -0.007\\
3400 & 1.0 & 1.2 & M & 1.12 & 0.28 &   70143 &   70071 &    5094 &    4783 & 0.681 & 0.682 &  0.001 & -0.000\\
3400 & 1.0 & 1.3 & M & 1.20 & 0.31 &   52156 &   52156 &    4217 &    3971 & 0.677 & 0.678 &  0.006 & -0.000\\
3400 & 1.0 & 1.4 & M & 1.30 & 0.28 &   38969 &   38969 &    3243 &    2928 & 0.668 & 0.670 &  0.007 &  0.001\\
3400 & 1.0 & 1.5 & M & 1.38 & 0.31 &   29518 &   29518 &    2691 &    2384 & 0.659 & 0.659 &  0.013 & -0.001\\
3400 & 1.0 & 1.6 & M & 1.48 & 0.29 &   22659 &   22659 &    2166 &    1894 & 0.646 & 0.649 &  0.012 & -0.002\\
3400 & 1.0 & 1.7 & M & 1.59 & 0.27 &   18175 &   18175 &    1753 &    1510 & 0.624 & 0.635 &  0.019 & -0.005\\
3400 & 1.0 & 1.8 & M & 1.67 & 0.30 &   13797 &   13797 &    1528 &    1227 & 0.620 & 0.605 &  0.013 & -0.006\\
3400 & 1.0 & 2.0 & M & 1.85 & 0.30 &    8644 &    8646 &    1132 &     818 & 0.552 & 0.574 &  0.033 & -0.006\\
3400 & 1.0 & 2.1 & M & 1.95 & 0.28 &    6885 &    6885 &     970 &     662 & 0.512 & 0.519 &  0.048 & -0.007\\
3400 & 1.0 & 2.2 & M & 2.03 & 0.32 &    5493 &    5493 &     876 &     541 & 0.467 & 0.470 &  0.059 & -0.008\\
3400 & 1.0 & 2.4 & M & 2.22 & 0.30 &    4206 &    4261 &     676 &     328 & 0.325 & 0.326 &  0.085 & -0.013\\
3700 & 0.4 & 1.2 & S & 1.14 & 0.25 &   16991 &  314052 &    4881 &    4436 & 0.698 & 0.687 &  0.039 & -0.000\\
3700 & 0.4 & 1.3 & S & 1.22 & 0.29 &   12092 &  243954 &    3962 &    3405 & 0.691 & 0.682 &  0.034 & -0.000\\
3700 & 0.4 & 1.4 & S & 1.32 & 0.29 &    8644 &  210585 &    3113 &    2540 & 0.690 & 0.674 &  0.046 & -0.002\\
3700 & 0.4 & 1.5 & S & 1.41 & 0.28 &    6247 &  181307 &    2526 &    1932 & 0.667 & 0.648 &  0.061 & -0.004\\
3700 & 0.4 & 1.8 & P & 1.68 & 0.34 &    3444 &  114628 &    1501 &     667 & 0.032 & 0.467 &  0.114 & -0.017\\
3700 & 0.5 & 1.5 & S & 1.41 & 0.26 &    8644 &  124021 &    2507 &    2054 & 0.678 & 0.656 &  0.038 & -0.004\\
3700 & 0.5 & 1.6 & P & 1.49 & 0.30 &    6382 &  102516 &    2111 &    1610 & 0.479 & 0.479 &  0.050 & -0.006\\
3700 & 0.5 & 1.7 & P & 1.59 & 0.29 &    4458 &   85148 &    1747 &    1196 & 0.391 & 0.391 &  0.063 & -0.007\\
3700 & 0.5 & 1.8 & P & 1.68 & 0.31 &    4206 &   70071 &    1494 &     850 & 0.251 & 0.251 &  0.070 & -0.006\\
3700 & 0.5 & 1.9 & S & 1.78 & 0.28 &    3438 &   57523 &    1263 &    1120 & 0.672 & 0.669 &  0.029 & -0.002\\
3700 & 0.6 & 1.2 & S & 1.12 & 0.27 &   29518 &  162623 &    4958 &    2885 & 0.689 & 0.686 &  0.008 &  0.001\\
3700 & 0.6 & 1.3 & S & 1.22 & 0.27 &   21253 &  133093 &    4010 &    3642 & 0.688 & 0.679 &  0.018 & -0.000\\
3700 & 0.6 & 1.4 & S & 1.31 & 0.27 &   15571 &  108190 &    3181 &    2796 & 0.686 & 0.672 &  0.018 & -0.001\\
3700 & 0.6 & 1.5 & S & 1.41 & 0.27 &   11552 &   87227 &    2479 &    2176 & 0.680 & 0.660 &  0.033 & -0.002\\
3700 & 0.6 & 1.6 & M & 1.49 & 0.30 &    8644 &   70071 &    2094 &    1699 & 0.677 & 0.645 &  0.035 & -0.004\\
3700 & 0.6 & 1.7 & P & 1.59 & 0.28 &    6504 &   56129 &    1765 &    1335 & 0.501 & 0.501 &  0.042 & -0.006\\
3700 & 0.6 & 1.8 & P & 1.68 & 0.28 &    4907 &   45016 &    1480 &    1010 & 0.403 & 0.403 &  0.052 & -0.007\\
3700 & 0.6 & 1.9 & P & 1.77 & 0.29 &    3734 &   36312 &    1275 &     753 & 0.283 & 0.283 &  0.060 & -0.006\\
3700 & 0.6 & 2.0 & P & 1.85 & 0.33 &    3461 &   29518 &    1133 &     556 & 0.120 & 0.134 &  0.094 & -0.012\\
3700 & 0.8 & 1.2 & S & 1.12 & 0.26 &   47280 &  102516 &    4972 &    3149 & 0.686 & 0.684 &  0.001 &  0.000\\
3700 & 0.8 & 1.3 & S & 1.21 & 0.29 &   34390 &   79923 &    4160 &    3844 & 0.682 & 0.678 &  0.003 & -0.000\\
3700 & 0.8 & 1.4 & S & 1.30 & 0.29 &   25450 &   61494 &    3261 &    2972 & 0.676 & 0.669 &  0.012 & -0.001\\
3700 & 0.8 & 1.5 & S & 1.40 & 0.28 &   19036 &   47262 &    2624 &    2303 & 0.658 & 0.657 &  0.009 & -0.002\\
3700 & 0.8 & 1.6 & M & 1.48 & 0.31 &   14553 &   36598 &    2137 &    1862 & 0.659 & 0.644 &  0.018 & -0.002\\
3700 & 0.8 & 1.7 & M & 1.58 & 0.29 &   11182 &   28644 &    1790 &    1466 & 0.589 & 0.589 &  0.022 & -0.005\\
3700 & 0.8 & 1.8 & P & 1.67 & 0.29 &    8644 &   22659 &    1512 &    1172 & 0.541 & 0.541 &  0.030 & -0.006\\
3700 & 0.8 & 1.9 & P & 1.75 & 0.32 &    6712 &   18089 &    1325 &     931 & 0.492 & 0.526 &  0.032 & -0.004\\
3700 & 0.8 & 2.0 & P & 1.84 & 0.32 &    5225 &   14553 &    1141 &     761 & 0.428 & 0.545 &  0.050 & -0.008\\
3700 & 0.8 & 2.1 & P & 1.94 & 0.30 &    4861 &   11784 &     985 &     619 & 0.345 & 0.463 &  0.071 & -0.008\\
3700 & 0.8 & 2.2 & P & 2.03 & 0.31 &    3956 &    9581 &     869 &     481 & 0.233 & 0.425 &  0.081 & -0.010\\
3700 & 0.8 & 2.3 & P & 2.12 & 0.32 &    3445 &    7811 &     772 &       1 & 0.050 & 0.050 &  0.153 &  0.060\\
3700 & 0.9 & 1.5 & S & 1.40 & 0.26 &   23930 &   36719 &    2528 &    2296 & 0.654 & 0.657 &  0.013 & -0.003\\
3700 & 0.9 & 1.6 & M & 1.48 & 0.29 &   18301 &   28286 &    2160 &    1871 & 0.650 & 0.647 &  0.014 & -0.003\\
3700 & 0.9 & 1.7 & M & 1.57 & 0.30 &   14150 &   22050 &    1808 &    1497 & 0.599 & 0.599 &  0.017 & -0.004\\
3700 & 0.9 & 1.8 & M & 1.67 & 0.28 &   11029 &   17369 &    1503 &    1204 & 0.572 & 0.572 &  0.026 & -0.006\\
3700 & 0.9 & 1.9 & P & 1.75 & 0.31 &    8644 &   13797 &    1319 &     958 & 0.541 & 0.582 &  0.026 & -0.004\\
3700 & 0.9 & 2.0 & P & 1.85 & 0.29 &    6801 &   11029 &    1121 &     777 & 0.492 & 0.558 &  0.040 & -0.006\\
3700 & 1.0 & 1.2 & M & 1.12 & 0.28 &   70143 &   70071 &    5094 &    4783 & 0.681 & 0.682 &  0.001 & -0.000\\
3700 & 1.0 & 1.3 & M & 1.20 & 0.31 &   52156 &   52156 &    4228 &    4030 & 0.675 & 0.675 &  0.001 &  0.000\\
3700 & 1.0 & 1.4 & M & 1.30 & 0.28 &   38969 &   38969 &    3236 &    2988 & 0.665 & 0.667 &  0.009 & -0.001\\
3700 & 1.0 & 1.5 & M & 1.40 & 0.26 &   29518 &   29518 &    2598 &    2268 & 0.650 & 0.651 &  0.010 & -0.000\\
3700 & 1.0 & 1.6 & M & 1.48 & 0.29 &   22659 &   22659 &    2178 &    1878 & 0.636 & 0.645 &  0.013 & -0.002\\
3700 & 1.0 & 1.7 & M & 1.58 & 0.28 &   18175 &   18175 &    1780 &    1490 & 0.618 & 0.625 &  0.018 & -0.005\\
3700 & 1.0 & 1.8 & M & 1.66 & 0.31 &   13797 &   13797 &    1542 &    1196 & 0.597 & 0.596 &  0.013 & -0.005\\
3700 & 1.0 & 1.9 & M & 1.76 & 0.29 &   10895 &   10895 &    1296 &     997 & 0.572 & 0.582 &  0.036 & -0.007\\
3700 & 1.0 & 2.0 & M & 1.84 & 0.32 &    8644 &    8646 &    1136 &     783 & 0.537 & 0.554 &  0.035 & -0.005\\
3700 & 1.0 & 2.1 & M & 1.94 & 0.30 &    6885 &    6885 &     983 &     641 & 0.493 & 0.492 &  0.053 & -0.008\\
3700 & 1.0 & 2.2 & M & 2.04 & 0.29 &    5493 &    5493 &     861 &     411 & 0.436 & 0.436 &  0.001 &  0.000\\
3700 & 1.0 & 2.4 & M & 2.20 & 0.34 &    4206 &    4261 &     695 &     305 & 0.281 & 0.281 &  0.098 & -0.015\\
4200 & 0.4 & 1.2 & S & 1.14 & 0.26 &   16991 &  314052 &    5119 &    4442 & 0.698 & 0.684 &  0.031 &  0.000\\
4200 & 0.4 & 1.3 & S & 1.23 & 0.26 &   12092 &  243954 &    3890 &    3272 & 0.689 & 0.676 &  0.037 & -0.001\\
4200 & 0.4 & 1.4 & S & 1.33 & 0.26 &    8644 &  210585 &    3052 &    2379 & 0.688 & 0.666 &  0.058 & -0.000\\
4200 & 0.6 & 1.2 & S & 1.12 & 0.29 &   29518 &  162623 &    5056 &    4728 & 0.688 & 0.684 &  0.028 &  0.000\\
4200 & 0.6 & 1.3 & S & 1.22 & 0.28 &   21253 &  133093 &    4050 &    3623 & 0.685 & 0.676 &  0.018 & -0.001\\
4200 & 0.6 & 1.4 & S & 1.31 & 0.28 &   15571 &  108190 &    3191 &    2755 & 0.683 & 0.666 &  0.022 & -0.002\\
4200 & 0.8 & 1.2 & S & 1.12 & 0.27 &   47280 &  102516 &    5003 &    4701 & 0.683 & 0.681 &  0.014 &  0.000\\
4200 & 0.8 & 1.3 & S & 1.20 & 0.30 &   34390 &   79923 &    4157 &    3820 & 0.679 & 0.675 &  0.005 & -0.000\\
4200 & 0.8 & 1.4 & S & 1.30 & 0.30 &   25450 &   61494 &    3273 &    2948 & 0.673 & 0.665 &  0.013 & -0.001\\
4200 & 1.0 & 1.2 & M & 1.12 & 0.28 &   70143 &   70071 &    5117 &    4757 & 0.678 & 0.679 &  0.001 &  0.000\\
4200 & 1.0 & 1.3 & M & 1.22 & 0.26 &   52156 &   52156 &    4049 &     270 & 0.670 & 0.671 &  0.004 &  0.000\\
4200 & 1.0 & 1.4 & M & 1.30 & 0.29 &   38969 &   38969 &    3256 &    2953 & 0.659 & 0.661 &  0.010 & -0.001\\
4200 & 1.0 & 2.2 & M & 2.04 & 0.30 &    5493 &    5493 &     864 &     454 & 0.386 & 0.386 &  0.083 & -0.011\\
4700 & 0.4 & 1.2 & S & 1.13 & 0.27 &   16991 &  314052 &    4934 &    4420 & 0.698 & 0.682 &  0.062 & -0.000\\
4700 & 0.4 & 1.3 & S & 1.23 & 0.28 &   12092 &  243954 &    3930 &    3213 & 0.687 & 0.671 &  0.042 & -0.001\\
4700 & 0.4 & 1.4 & S & 1.32 & 0.28 &    8644 &  210585 &    3103 &    2310 & 0.686 & 0.669 &  0.058 & -0.003\\
4700 & 0.6 & 1.2 & S & 1.13 & 0.25 &   29518 &  162623 &    5176 &    4571 & 0.684 & 0.681 &  0.006 &  0.001\\
4700 & 0.6 & 1.3 & S & 1.21 & 0.28 &   21253 &  133093 &    4072 &      12 & 0.685 & 0.673 &  0.007 &  0.000\\
4700 & 0.6 & 1.4 & S & 1.31 & 0.29 &   15571 &  108190 &    3215 &    2720 & 0.680 & 0.660 &  0.025 & -0.002\\
4700 & 0.8 & 1.2 & S & 1.12 & 0.27 &   47280 &  102516 &    5023 &    4810 & 0.680 & 0.678 &  0.021 &  0.000\\
4700 & 0.8 & 1.3 & S & 1.20 & 0.30 &   34390 &   79923 &     486 &      14 & 0.676 & 0.671 &  0.000 &  0.000\\
4700 & 0.8 & 1.4 & S & 1.29 & 0.31 &   25450 &   61494 &    3299 &    3009 & 0.670 & 0.659 &  0.035 & -0.006\\
4700 & 1.0 & 1.2 & M & 1.11 & 0.28 &   70143 &   70071 &    5140 &    4725 & 0.675 & 0.676 &  0.000 & -0.000\\
4700 & 1.0 & 1.3 & M & 1.21 & 0.27 &   52156 &   52156 &    4059 &    3848 & 0.667 & 0.667 &  0.006 & -0.000\\
4700 & 1.0 & 1.4 & M & 1.30 & 0.30 &   38969 &   38969 &    3275 &    2922 & 0.652 & 0.655 &  0.011 & -0.001\\
4700 & 1.0 & 2.2 & M & 2.03 & 0.32 &    5493 &    5493 &     877 &     431 & 0.329 & 0.329 &  0.100 & -0.015\\
5200 & 0.4 & 1.2 & S & 1.13 & 0.29 &   16991 &  314052 &    5253 &    4375 & 0.698 & 0.680 &  0.044 &  0.001\\
5200 & 0.4 & 1.3 & S & 1.22 & 0.29 &   12092 &  243954 &    3955 &    3166 & 0.687 & 0.669 &  0.050 & -0.001\\
5200 & 0.4 & 1.4 & M & 1.33 & 0.26 &    8644 &  210585 &    3061 &    2251 & 0.447 & 0.421 &  0.067 & -0.004\\
5200 & 0.6 & 1.2 & S & 1.13 & 0.25 &   29518 &  162623 &    5250 &    4687 & 0.684 & 0.679 &  0.020 &  0.000\\
5200 & 0.6 & 1.3 & S & 1.22 & 0.26 &   21253 &  133093 &    3970 &    3471 & 0.682 & 0.668 &  0.020 & -0.001\\
5200 & 0.6 & 1.4 & S & 1.31 & 0.26 &   15571 &  108190 &    3153 &    2628 & 0.679 & 0.653 &  0.029 & -0.002\\
5200 & 0.8 & 1.2 & S & 1.12 & 0.27 &   47280 &  102516 &    5048 &    2917 & 0.680 & 0.676 &  0.000 &  0.000\\
5200 & 0.8 & 1.3 & S & 1.22 & 0.26 &   34390 &   79923 &      75 &    3614 & 0.672 & 0.666 &  0.277 & -0.006\\
5200 & 0.8 & 1.4 & S & 1.29 & 0.31 &   25450 &   61494 &    3311 &    2888 & 0.667 & 0.653 &  0.016 & -0.002\\
5200 & 1.0 & 1.2 & M & 1.11 & 0.29 &   70143 &   70071 &    5162 &    4703 & 0.672 & 0.673 &  0.001 & -0.000\\
5200 & 1.0 & 1.3 & M & 1.21 & 0.27 &   52156 &   52156 &    4073 &    3689 & 0.661 & 0.662 &  0.009 & -0.001\\
5200 & 1.0 & 1.4 & M & 1.30 & 0.30 &   38969 &   38969 &    3286 &    2910 & 0.649 & 0.652 &  0.012 & -0.001\\
5200 & 1.0 & 2.2 & M & 2.02 & 0.32 &    5493 &    5493 &     880 &     379 & 0.263 & 0.263 &  0.125 & -0.018\\
5700 & 0.4 & 1.2 & S & 1.13 & 0.30 &   16991 &  314052 &    4884 &    3914 & 0.698 & 0.677 &  0.009 & -0.002\\
5700 & 0.4 & 1.3 & S & 1.23 & 0.26 &   12092 &  243954 &    3897 &    3022 & 0.684 & 0.661 &  0.053 & -0.001\\
5700 & 0.6 & 1.2 & S & 1.13 & 0.26 &   29518 &  162623 &    4626 &    4641 & 0.681 & 0.675 &  0.081 & -0.001\\
5700 & 0.6 & 1.3 & S & 1.22 & 0.26 &   21253 &  133093 &    3551 &    3463 & 0.679 & 0.664 &  0.125 &  0.005\\
5700 & 0.6 & 1.4 & S & 1.31 & 0.27 &   15571 &  108190 &    3174 &    2592 & 0.676 & 0.647 &  0.032 & -0.003\\
5700 & 0.8 & 1.2 & S & 1.12 & 0.28 &   47280 &  102516 &    5068 &    4926 & 0.677 & 0.673 &  0.041 & -0.001\\
5700 & 0.8 & 1.3 & S & 1.22 & 0.26 &   34390 &   79923 &    4040 &      13 & 0.669 & 0.662 &  0.004 &  0.000\\
5700 & 0.8 & 1.4 & S & 1.31 & 0.27 &   25450 &   61494 &    3197 &    2742 & 0.662 & 0.648 &  0.014 & -0.002\\
5700 & 1.0 & 1.2 & M & 1.11 & 0.29 &   70143 &   70071 &    5183 &    4664 & 0.669 & 0.671 &  0.000 & -0.000\\
5700 & 1.0 & 1.3 & M & 1.21 & 0.27 &   52156 &   52156 &    4127 &    3669 & 0.658 & 0.659 &  0.004 & -0.000\\
5700 & 1.0 & 1.4 & M & 1.29 & 0.31 &   38969 &   38969 &    3304 &    2876 & 0.643 & 0.646 &  0.014 & -0.001\\

\end{longtable}

}
}

\addtocounter{table}{1}
\onltab{\arabic{table}}{
\longtab{\arabic{table}}{

\begin{longtable}{rrrrrrrrrrrrrr}
\caption{As Table \ref{tab:model_table} for the $Z=0.001$ models.}\label{tab:model_z001_table}\\
\hline\hline
$t$ & $q$ & $M$ & Case & $M_\mathrm{\remnantabbr}$ & $\frac{(1+q)^2}{q}\phi$ &
$\tau_\mathrm{ms, 1}$ & $\tau_\mathrm{ms, 2}$ & $\tau_\mathrm{ms}$ &
$t_\mathrm{ms}$ & $X_{\mathrm{c},0}$ & $X_{\mathrm{c,zms}}$ & $\Delta
\logten L$ & $\Delta \logten T_\mathrm{eff}$\\
\hline
\endfirsthead

\caption{cont.}\\
\hline\hline
$t$ & $q$ & $M$ & Case & $M_\mathrm{\remnantabbr}$ & $\frac{(1+q)^2}{q}\phi$ &
$\tau_\mathrm{ms, 1}$ & $\tau_\mathrm{ms, 2}$ & $\tau_\mathrm{ms}$ &
$t_\mathrm{ms}$ & $X_{\mathrm{c},0}$ & $X_{\mathrm{c,zms}}$ & $\Delta
\logten L$ & $\Delta \logten T_\mathrm{eff}$\\
\hline
\endhead

\hline
\endfoot
8000 & 0.4 & 0.8 & S & 0.76 & 0.25 &   43810 &  695587 &   14678 &   13000 & 0.753 & 0.751 &  0.018 &  0.001\\
8000 & 0.4 & 0.9 & S & 0.85 & 0.25 &   27922 &  544208 &    9331 &    7683 & 0.752 & 0.719 &  0.032 &  0.002\\
8000 & 0.4 & 1.0 & S & 0.95 & 0.25 &   18612 &  418946 &    6214 &    4645 & 0.752 & 0.701 &  0.062 &  0.006\\
8000 & 0.4 & 1.1 & P & 1.04 & 0.27 &   12867 &  313659 &    4353 &    2459 & 0.408 & 0.403 &  0.066 &  0.010\\
8000 & 0.4 & 1.2 & P & 1.13 & 0.29 &    9181 &  221923 &    3150 &       3 & 0.210 & 0.210 &  0.000 &  0.089\\
8000 & 0.6 & 0.8 & S & 0.76 & 0.24 &   72663 &  364938 &   14963 &   13905 & 0.752 & 0.752 &  0.015 &  0.001\\
8000 & 0.6 & 0.9 & S & 0.85 & 0.24 &   46528 &  236659 &    9496 &    8484 & 0.751 & 0.746 &  0.016 &  0.001\\
8000 & 0.6 & 1.0 & S & 0.93 & 0.28 &   31103 &  178559 &    6579 &    5470 & 0.730 & 0.722 &  0.019 &  0.001\\
8000 & 0.6 & 1.1 & M & 1.03 & 0.27 &   21571 &  138218 &    4513 &    2827 & 0.573 & 0.541 &  0.030 &  0.001\\
8000 & 0.6 & 1.2 & P & 1.12 & 0.28 &   15409 &  105374 &    3241 &    1986 & 0.485 & 0.485 &  0.027 &  0.004\\
8000 & 0.6 & 1.3 & P & 1.22 & 0.28 &   11298 &   79734 &    2354 &     935 & 0.348 & 0.350 &  0.028 &  0.006\\
8000 & 0.8 & 0.8 & S & 0.75 & 0.25 &  109778 &  202645 &   15353 &   14290 & 0.735 & 0.733 &  0.004 &  0.000\\
8000 & 0.8 & 0.9 & S & 0.85 & 0.24 &   72663 &  150804 &    9633 &    8725 & 0.727 & 0.727 &  0.004 &  0.000\\
8000 & 0.8 & 1.0 & S & 0.94 & 0.25 &   48790 &  109778 &    6487 &    5614 & 0.720 & 0.713 &  0.006 &  0.000\\
8000 & 0.8 & 1.2 & P & 1.12 & 0.26 &   24284 &   57012 &    3215 &    2305 & 0.592 & 0.592 &  0.011 &  0.001\\
8000 & 0.8 & 1.3 & P & 1.21 & 0.30 &   17831 &   41997 &    2441 &    1493 & 0.526 & 0.527 &  0.025 &  0.002\\
8000 & 1.0 & 0.8 & M & 0.75 & 0.26 &  150804 &  150804 &   15550 &   14471 & 0.729 & 0.729 &  0.003 &  0.000\\
8000 & 1.0 & 0.9 & M & 0.84 & 0.27 &  105374 &  105374 &    9911 &    9016 & 0.720 & 0.720 &  0.003 &  0.000\\
8000 & 1.0 & 1.0 & M & 0.93 & 0.27 &   72663 &   72663 &    6603 &    5773 & 0.705 & 0.705 &  0.003 &  0.000\\
8000 & 1.0 & 1.1 & M & 1.03 & 0.26 &   50701 &   50701 &    4509 &    3695 & 0.682 & 0.683 &  0.008 &  0.001\\
8000 & 1.0 & 1.2 & M & 1.12 & 0.26 &   36358 &   36358 &    3218 &    2388 & 0.649 & 0.651 &  0.007 &  0.001\\
8000 & 1.0 & 1.3 & M & 1.21 & 0.29 &   26764 &   26764 &    2425 &    1596 & 0.612 & 0.612 &  0.024 &  0.001\\
9500 & 0.4 & 0.8 & S & 0.76 & 0.26 &   43810 &  695587 &   14847 &   12884 & 0.752 & 0.752 &  0.023 &  0.002\\
9500 & 0.4 & 0.9 & S & 0.85 & 0.27 &   27922 &  544208 &    9479 &    7438 & 0.751 & 0.709 &  0.037 &  0.002\\
9500 & 0.4 & 1.0 & S & 0.94 & 0.28 &   18612 &  418946 &    6367 &    4278 & 0.751 & 0.680 &  0.052 &  0.006\\
9500 & 0.4 & 1.1 & P & 1.04 & 0.28 &   12867 &  313659 &    4403 &    2057 & 0.313 & 0.313 &  0.069 &  0.011\\
9500 & 0.4 & 1.2 & P & 1.13 & 0.29 &    9181 &  221923 &    3151 &     235 & 0.047 & 0.069 &  0.079 &  0.017\\
9500 & 0.6 & 0.8 & S & 0.75 & 0.24 &   72663 &  364938 &   15051 &   13767 & 0.751 & 0.750 &  0.015 &  0.001\\
9500 & 0.6 & 0.9 & S & 0.85 & 0.25 &   46528 &  236659 &    1009 &    8353 & 0.751 & 0.729 &  0.500 &  0.025\\
9500 & 0.6 & 1.0 & S & 0.94 & 0.25 &   31103 &  178559 &    6412 &    5105 & 0.723 & 0.705 &  0.022 &  0.002\\
9500 & 0.6 & 1.1 & P & 1.03 & 0.26 &   21571 &  138218 &    4469 &    3073 & 0.528 & 0.528 &  0.028 &  0.003\\
9500 & 0.6 & 1.2 & P & 1.12 & 0.28 &   15409 &  105374 &    3247 &    1695 & 0.414 & 0.415 &  0.031 &  0.004\\
9500 & 0.6 & 1.3 & P & 1.21 & 0.30 &   11298 &   79734 &    2420 &     621 & 0.235 & 0.235 &  0.032 &  0.007\\
9500 & 0.8 & 0.8 & S & 0.75 & 0.26 &  109778 &  202645 &   15447 &   14171 & 0.729 & 0.728 &  0.006 &  0.000\\
9500 & 0.8 & 0.9 & S & 0.84 & 0.25 &   72663 &  150804 &    9702 &    8612 & 0.722 & 0.720 &  0.005 &  0.000\\
9500 & 0.8 & 1.1 & M & 1.02 & 0.28 &   33896 &   78905 &    4632 &    3545 & 0.620 & 0.631 &  0.010 &  0.002\\
9500 & 0.8 & 1.2 & P & 1.12 & 0.28 &   24284 &   57012 &    3279 &    2173 & 0.557 & 0.557 &  0.014 &  0.002\\
9500 & 0.8 & 1.3 & P & 1.20 & 0.30 &   17831 &   41997 &    2336 &    1285 & 0.470 & 0.470 &  0.050 &  0.003\\
9500 & 1.0 & 0.8 & M & 0.75 & 0.27 &  150804 &  150804 &   15638 &   14353 & 0.724 & 0.724 &  0.002 &  0.000\\
9500 & 1.0 & 0.9 & M & 0.84 & 0.27 &  105374 &  105374 &    9974 &    8906 & 0.713 & 0.713 &  0.003 &  0.000\\
9500 & 1.0 & 1.0 & M & 0.93 & 0.28 &   72663 &   72663 &    6663 &    5653 & 0.695 & 0.695 &  0.004 &  0.000\\
9500 & 1.0 & 1.1 & M & 1.03 & 0.27 &   50701 &   50701 &    4584 &    3588 & 0.667 & 0.667 &  0.007 &  0.001\\
9500 & 1.0 & 1.2 & M & 1.12 & 0.27 &   36358 &   36358 &    3271 &    2264 & 0.627 & 0.631 &  0.010 &  0.001\\
9500 & 1.0 & 1.3 & M & 1.21 & 0.26 &   26764 &   26764 &    2368 &    1410 & 0.574 & 0.579 &  0.032 &  0.001\\
11000 & 0.4 & 0.8 & S & 0.76 & 0.28 &   43810 &  695587 &   14999 &   12605 & 0.751 & 0.746 &  0.028 &  0.002\\
11000 & 0.4 & 0.9 & S & 0.85 & 0.25 &   27922 &  544208 &    9333 &    7069 & 0.751 & 0.710 &  0.054 &  0.005\\
11000 & 0.4 & 1.0 & P & 0.94 & 0.28 &   18612 &  418946 &    6329 &    3858 & 0.420 & 0.420 &  0.059 &  0.006\\
11000 & 0.4 & 1.1 & P & 1.03 & 0.30 &   12867 &  313659 &    4484 &    1377 & 0.211 & 0.211 &  0.036 &  0.011\\
11000 & 0.6 & 0.8 & S & 0.75 & 0.25 &   72663 &  364938 &   15139 &   13640 & 0.750 & 0.749 &  0.015 &  0.001\\
11000 & 0.6 & 0.9 & S & 0.85 & 0.26 &   46528 &  236659 &    9674 &    8250 & 0.750 & 0.737 &  0.023 &  0.002\\
11000 & 0.6 & 1.0 & M & 0.94 & 0.27 &   31103 &  178559 &    6499 &    4849 & 0.696 & 0.536 &  0.010 &  0.002\\
11000 & 0.6 & 1.1 & P & 1.03 & 0.25 &   21571 &  138218 &    4434 &    2770 & 0.480 & 0.480 &  0.030 &  0.003\\
11000 & 0.6 & 1.2 & P & 1.12 & 0.29 &   15409 &  105374 &    3264 &    1322 & 0.336 & 0.336 &  0.026 &  0.005\\
11000 & 0.6 & 1.3 & P & 1.21 & 0.29 &   11298 &   79734 &    2383 &     306 & 0.104 & 0.134 &  0.071 &  0.009\\
11000 & 0.8 & 0.8 & S & 0.75 & 0.26 &  109778 &  202645 &   15514 &   14031 & 0.724 & 0.724 &  0.005 &  0.000\\
11000 & 0.8 & 0.9 & S & 0.84 & 0.27 &   72663 &  150804 &    9928 &    8636 & 0.717 & 0.713 &  0.005 &  0.000\\
11000 & 0.8 & 1.1 & P & 1.02 & 0.30 &   33896 &   78905 &    4696 &    3407 & 0.596 & 0.596 &  0.012 &  0.002\\
11000 & 0.8 & 1.2 & P & 1.12 & 0.27 &   24284 &   57012 &    3257 &    1940 & 0.516 & 0.516 &  0.015 &  0.002\\
11000 & 0.8 & 1.3 & P & 1.21 & 0.28 &   17831 &   41997 &    2401 &    1026 & 0.407 & 0.407 &  0.019 &  0.002\\
11000 & 1.0 & 0.8 & M & 0.75 & 0.27 &  150804 &  150804 &   15694 &   14943 & 0.720 & 0.720 &  0.000 &  0.000\\
11000 & 1.0 & 0.9 & M & 0.84 & 0.28 &  105374 &  105374 &   10036 &     751 & 0.706 & 0.706 &  0.000 &  0.000\\
11000 & 1.0 & 1.0 & M & 0.93 & 0.28 &   72663 &   72663 &    6723 &    5559 & 0.685 & 0.685 &  0.003 &  0.000\\
11000 & 1.0 & 1.1 & M & 1.02 & 0.28 &   50701 &   50701 &    4639 &    3473 & 0.653 & 0.653 &  0.007 &  0.001\\
11000 & 1.0 & 1.2 & M & 1.11 & 0.29 &   36358 &   36358 &    3332 &    2125 & 0.606 & 0.606 &  0.011 &  0.001\\
11000 & 1.0 & 1.3 & M & 1.21 & 0.29 &   26764 &   26764 &    2321 &    1175 & 0.541 & 0.545 &  0.025 &  0.001\\
12500 & 0.4 & 0.9 & S & 0.85 & 0.27 &   27922 &  544208 &    9492 &    6505 & 0.750 & 0.690 &  0.014 &  0.004\\
12500 & 0.4 & 1.0 & P & 0.94 & 0.27 &   18612 &  418946 &    6304 &    3365 & 0.355 & 0.355 &  0.067 &  0.007\\
12500 & 0.4 & 1.1 & P & 1.03 & 0.29 &   12867 &  313659 &    4440 &     492 & 0.095 & 0.096 &  0.047 &  0.016\\
12500 & 0.6 & 0.8 & S & 0.75 & 0.26 &   72663 &  364938 &   15230 &   13508 & 0.750 & 0.747 &  0.017 &  0.001\\
12500 & 0.6 & 0.9 & S & 0.85 & 0.24 &   46528 &  236659 &    9475 &    7857 & 0.748 & 0.718 &  0.025 &  0.002\\
12500 & 0.6 & 1.0 & M & 0.94 & 0.25 &   31103 &  178559 &    6399 &    4582 & 0.546 & 0.546 &  0.027 &  0.002\\
12500 & 0.6 & 1.1 & P & 1.03 & 0.28 &   21571 &  138218 &    4537 &    2526 & 0.433 & 0.433 &  0.029 &  0.004\\
12500 & 0.6 & 1.2 & P & 1.12 & 0.30 &   15409 &  105374 &    3301 &     918 & 0.249 & 0.249 &  0.022 &  0.007\\
12500 & 0.8 & 0.8 & S & 0.75 & 0.27 &  109778 &  202645 &   15605 &   13908 & 0.720 & 0.719 &  0.007 &  0.000\\
12500 & 0.8 & 0.9 & S & 0.84 & 0.28 &   72663 &  150804 &   10004 &    8516 & 0.712 & 0.706 &  0.007 &  0.000\\
12500 & 0.8 & 1.0 & M & 0.94 & 0.26 &   48790 &  109778 &    6515 &    5108 & 0.630 & 0.633 &  0.008 &  0.001\\
12500 & 0.8 & 1.1 & P & 1.03 & 0.27 &   33896 &   78905 &    4552 &    3115 & 0.567 & 0.567 &  0.013 &  0.002\\
12500 & 0.8 & 1.2 & P & 1.11 & 0.29 &   24284 &   57012 &    3326 &    1796 & 0.476 & 0.476 &  0.016 &  0.002\\
12500 & 0.8 & 1.3 & P & 1.21 & 0.27 &   17831 &   41997 &    2366 &     817 & 0.338 & 0.338 &  0.027 &  0.002\\
12500 & 1.0 & 0.8 & M & 0.75 & 0.23 &  150804 &  150804 &   15091 &   13462 & 0.712 & 0.711 &  0.003 &  0.000\\
12500 & 1.0 & 0.9 & M & 0.84 & 0.28 &  105374 &  105374 &   10100 &    8680 & 0.699 & 0.699 &  0.004 &  0.000\\
12500 & 1.0 & 1.0 & M & 0.94 & 0.25 &   72663 &   72663 &    6464 &    5187 & 0.672 & 0.672 &  0.003 &  0.000\\
12500 & 1.0 & 1.1 & M & 1.02 & 0.29 &   50701 &   50701 &    4705 &    3866 & 0.639 & 0.639 &  0.003 &  0.001\\
12500 & 1.0 & 1.2 & M & 1.12 & 0.26 &   36358 &   36358 &    3235 &    1911 & 0.576 & 0.580 &  0.012 &  0.001\\
12500 & 1.0 & 1.3 & M & 1.21 & 0.27 &   26764 &   26764 &    2376 &    1126 & 0.502 & 0.507 &  0.049 &  0.002\\

\end{longtable}

}
}

\end{document}